\documentclass{llncs}
\usepackage{url}
\usepackage{graphicx}
\usepackage{cite}

\usepackage{subfigure}
\usepackage{algorithm}
\usepackage{algpseudocode}
\usepackage{makecell}
\usepackage{amssymb}
\usepackage{amsmath}

\usepackage{threeparttable}
\newcommand{\algoname}[1]{\textnormal{\textsc{#1}}}

\begin{document}

\title{On Counting Triangles through Edge Sampling in Large Dynamic Graphs}
\author{Guyue Han and Harish Sethu}
\institute{Department of ECE, Drexel University\\
Philadelphia, PA 19104\\
Email: \{guyue.han, sethu\}@drexel.edu}

\maketitle
\thispagestyle{empty}
~\vskip 0.5in
\begin{abstract}

Traditional frameworks for dynamic graphs have relied on processing only the stream of edges added into or deleted from an evolving graph, but not any additional related information such as the degrees or neighbor lists of nodes incident to the edges. In this paper, we propose a new edge sampling framework for big-graph analytics in dynamic graphs which enhances the traditional model by enabling the use of additional related information. To demonstrate the advantages of this framework, we present a new sampling algorithm, called {\em Edge Sample and Discard} (\algoname{esd}). It generates an unbiased estimate of the total number of triangles, which can be continuously updated in response to both edge additions and deletions. We provide a comparative analysis of the performance of \algoname{esd} against two current state-of-the-art algorithms in terms of accuracy and complexity. The results of the experiments performed on real graphs show that, with the help of the neighborhood information of the sampled edges, the accuracy achieved by our algorithm is substantially better. We also characterize the impact of properties of the graph on the performance of our algorithm by testing on several Barab$\mathrm{\acute a}$si-Albert graphs.



\end{abstract}

\newpage
\section{Introduction}
\label{sec:intro}
Given the rising significance of social networks in our society,
the analysis of their structural properties and the principles guiding their evolution and dynamics have attracted tremendous interest from researchers, sociologists and marketeers \cite{TirHal2015,FouVan2010,BerHen2011}. Social networks can be modeled as graphs with
nodes representing users and edges representing the interactions between the users; the study of social networks, therefore, usually translates into a study of extremely large graphs. 

In the real world, social networking services (social networking sites or social media), such as facebook, twitter and wechat, offer prominent examples of fully dynamic graphs.
A typical representation of a dynamic graph consists of two components: a
connected graph and an edge stream. The stream indicates the addition
of a new edge or the deletion of an existing edge from the graph. 
Social networking service (SNS) providers need to maintain and update their datasets in a real-time fashion. 
Moreover, service providers may perform various types of analytics on their graph datasets, such as distinguishing different communities, detecting anomalies or spam, and finding the nodes with high betweenness centrality. Real-time graph analytics has the ability to discover important information which can help SNS providers improve existing services, develop new ones, detect anomalous conditions and respond rapidly to resource management concerns.

One of the key structural properties of interest in social graphs is the triangle, the simplest of graph motifs. 
The number of triangles is used as one of the signatures of social roles in online discussion
groups \cite{WelGle2007}. The distribution of triangles is a relevant property for spam detection in social networks
\cite{BecBol2010}. 
The real-time estimation of the number of triangles helps monitor the evolution of the community structure of the graph. The global clustering coefficient can be easily tracked by the changes in the number of triangles and can tell us whether the network is becoming tightly connected, or decentralized \cite{New2003,BerHen2011}. Moreover, a dramatic growth or reduction in the number of triangles in a short time can reflect abnormal behaviors.

In this paper, we develop a new low-cost sampling algorithm which monitors the edge stream of an evolving graph and is able to, at any instant, provide the current real-time estimate of the number of triangles in it. The goal is for our algorithm to also be adaptable to the case of a static graph. 


A dynamic graph, such as one representing a social network, is described by a sequence of edge addition and deletion operations occurring over time. The following tasks may be involved in the management of the graph:
\begin{itemize}
\item {\em Maintenance of the graph datasets.} When a user becomes another user's follower or when a user removes some infrequent contacts from his/her friend list, the system needs to perform edge addition or deletion over the dataset. In addition, the system need to update the lists of neighbors of the users accordingly.
\item {\em Graph analytics.} Besides the task of maintaining the graph dataset mentioned above, some SNS providers may analyze their social networks quantitatively and qualitatively for better understanding of the networks and improving the existing services.
\end{itemize}
The fact that a typical dynamic graph-based system needs to maintain the graph dataset as described in the first of the two tasks above whether or not the second task of graph analytics is performed, suggests a framework where the minimum available information for graph analytics is all of the information obtained from the first task. In a real scenario, since graph datasets are constantly maintained anyway, it is unnecessary to restrict graph analytics to use only the information from the edge stream but without use of any information about other graph characteristics related to those edges. 

\subsection{A Framework for Graph Analytics}
\label{sec:framework}
\begin{figure}[!t]
\centering{
       \includegraphics[width=2.8in]{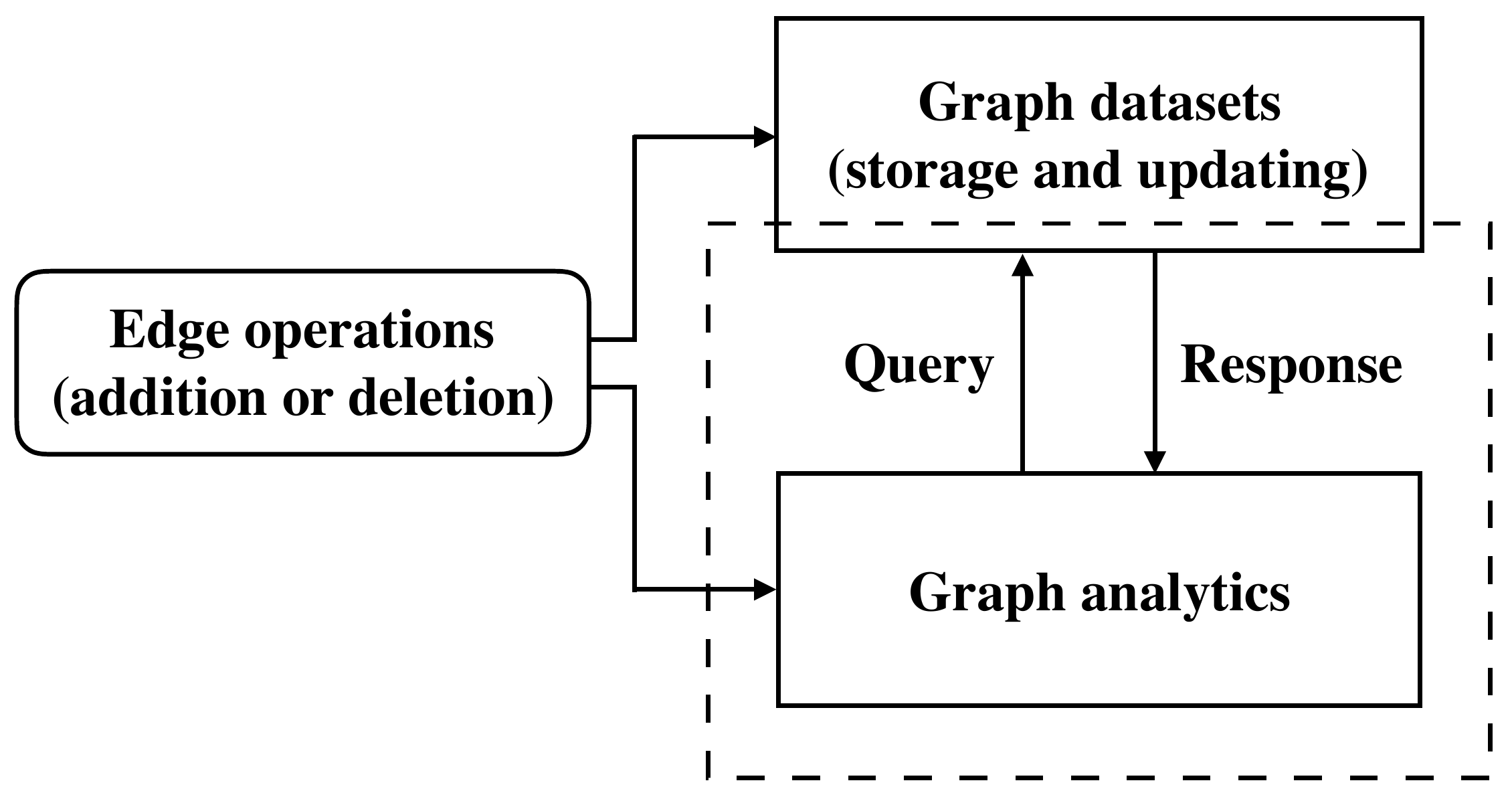}
       
    }
   \caption{A framework of graph analytics in a dynamic system. }\label{fig:system_outline} 
\end{figure} 

Figure \ref{fig:system_outline} shows the outline of our framework for
performing graph analytics in a dynamic system. Each time a new edge
operation happens, the system maintains the graph dataset by adding or
deleting the edge according to the operation, and updates the corresponding information which is determined by the service provided. This comprises the normal functioning of a dynamic graph-based system regardless of whether graph analytics is taken into consideration or not. Note that graph datasets have to be stored somewhere, typically on a server, and can be queried for graph analytics. In almost all real contexts, after all, we would not be counting the triangles in a graph without the graph existing in storage somewhere. Our framework, by assuming the existence of a stored graph dataset and allowing queries on it, achieves a better approximation of the reality of graph-based applications and networks.

For graph analytics, each new edge operation represents the addition or the deletion of an edge. There are streaming edge algorithms for various analytical purposes. Real-time algorithms for graph analytics will typically allow only a single pass over the edge stream. Based on the design of the algorithm, certain queries can be made of the server asking for additional information when processing a new edge. The availability of this additional information, via queries, is not an additional burden on the server since this information is often already updated and maintained by the service providers for offering necessary services. 

In our framework for graph analytics, we assume the ability to process an edge stream (one pass) and also the ability to query the graph dataset server for information on the neighborhood of an edge. The dotted line in the figure frames the real-time graph analytics in a realistic scenario. In this framework, each new edge operation is treated as an arriving edge involving either an addition of an edge or the deletion of an edge. Edges are sampled independently with a certain probability. If an edge is sampled, queries are sent to the server asking for the information on the neighborhood of the two incident nodes of the sampled edge in order to update the estimators used in graph analytics.

We use the above framework to design a new sampling algorithm to keep
a running count of the number of triangles in a dynamic graph. The
estimation is enabled by the use of neighborhood information. Since the list of neighbors of a user is necessary information for a social network's server to maintain and update anyway, little extra computational/memory costs are expended for querying this additional information.

\subsection{Contributions}
Based on the new framework/model of a dynamic graph system, we propose a new edge sampling algorithm, called {\em Edge Sample and Discard} (\algoname{esd}), which returns an estimate 
of the total number of triangles in a large dynamic graph by sampling only a
tiny fraction of the edges in the graph. For each sampled edge, it
samples the presence of triangles at the end points of the edge to
update its estimates and then discards the edge (as opposed to holding
the edge in memory in a subgraph sample as in
\cite{AhmDuf2014,TsoKan2009,JhaSes2015space}). The algorithm works on
fully dynamic graphs where both edge deletions and additions are
considered, and can be readily applied to static graphs as well. A preliminary version of this work appeared in \cite{HanSet2017}. 

In Section \ref{sec:rationale}, we introduce the graph model more formally and
present the \algoname{esd} algorithm. We show that the total number of
triangles can be estimated from the probabilities with which we sample
an edge and one of its neighbor nodes, and whether the sampled edge
and the node form a triangle. 

Section \ref{sec:estvar} presents a theoretical analysis of the
algorithm and proves that its estimate is an unbiased one. We also
derive a bound on the variance of our estimate and draw 
implications from it. We show that our sampling algorithm for
an evolving graph can be readily modified to apply to static graphs.

Section \ref{sec:simulation} provides a comparative analysis of \algoname{esd} against two edge sampling
algorithms, \algoname{doulion} \cite{TsoKan2009} and
\algoname{tri\`est} \cite{DBLP:journals/corr/SteEpa2016} which also
can handle both edge additions and deletions, and provide a real-time
tracking of the number of triangles. Note that the framework assumed by both \algoname{doulion} and \algoname{tri\`est} are slightly different from the one assumed by our algorithm. They are designed assuming traditional streaming edge framework and do not consider the possibility of using additional information obtained by querying the stored graph dataset on the server. 
We use several real network graphs with millions of edges to create streams of dynamic graphs. Based on tests on these graphs, we show that, with the use of the neighborhood information of the sampled edges, the accuracy achieved by \algoname{esd} is at least one order of magnitude better than the accuracies achieved by \algoname{doulion} and \algoname{tri\`est}, while the extra cost of querying for additional information is relatively small. These costs and the associated trade-offs are described in Section \ref{sec:simulation}. 

Section \ref{sec:simulation} also evaluates our algorithm on real dynamic graphs. The tests show that our algorithm can generate accurate  estimates of the number of triangles in real dynamic graphs. Moreover, we present the influence of the total number of triangles and the clustering coefficient on the accuracy of the estimate of \algoname{esd} by performing simulations on Barab$\mathrm{\acute a}$si-Albert (BA) graphs. The accuracy achieved by our algorithm is better on graphs with a larger number of triangles or a larger global clustering coefficient.

Section \ref{sec:conclusion} concludes the paper.

\subsection{Related Work}
Triadic properties such as triangle counts and the global clustering
coefficient have been widely studied \cite{ChaFal2006,WasFau1994}. Some
early works use enumeration or matrix multiplication to compute the
exact number of triangles in the graph
\cite{SchWag2005,Lat2008,CopWin1987,ChiNis1985}. Alon {\em et al.}
\cite{AloYus1997} propose the theoretically fastest exact triangle
counting algorithm, which is based on fast matrix multiplication and
runs in $O(|E|^{1.41})$ time. But it has a high space complexity of
$O(|V|^2)$ which renders it largely infeasible for extremely large
graphs. Besides, in many cases, the exact answer is not necessary and
an approximation is sufficient. Therefore, well-performing
approximation methods, which achieve fast runtime and small memory
footprint, have attracted tremendous interest.

Eigenvalue-based methods are one class of algorithms used to
approximate the global and local number of triangles in the graph
\cite{Avr2010,TsoCha2008}. They use the interesting property that 
the total number of triangles in an undirected graph is $1/6$ of the
sum of the cubes of the eigenvalues of its adjacency matrix. 
But the computation of matrix multiplication is still very expensive and they
only work on static graphs. Hardiman {\em et al.} \cite{Harkat2013}
presents a method based on a random walk which is capable of
estimating both the global and the average clustering coefficient by
testing the connectivity of each node in the random walk after the
mixing time is reached. 

Most of the studies on triangle counting use the graph stream model,
where a graph is treated as a stream of edges. Two algorithms for
approximating the local number of triangles in both directed and
undirected graph are presented in \cite{BecBol2010}. 
These two algorithms both require multiple passes over the edge
stream and only work on static graphs. 

Another class of algorithms uses an edge sparsification approach 
based on a certain selecting probability to decide whether an edge
should be sampled \cite{AhmDuf2014,TsoKan2009,LimKan2015}. In
\cite{KolMil2012}, a hybrid approach is used which combines edge
sparsification with degree-based vertex partitioning.
In all of these
algorithms, the sample size is not fixed. 
Algorithms bases on reservoir sampling, which use a fixed amount of space for estimating the triadic properties, are described in \cite{JhaSes2015space,Shin2017}. However,
except for the method presented in \cite{TsoKan2009}, none of these
algorithms can handle edge deletions in an evolving graph; they work
on static graphs and on dynamic graphs with edge additions
but not edge deletions.
 
One approach which works on fully dynamic graphs where both edge
additions and deletions are allowed is presented in
\cite{KutPag2014}. It combines the sampling of vertex triples
algorithm in \cite{BurFra2006} and monochromatic sampling method in
\cite{PagTso2012}. The algorithm first performs monochromatic sampling
on the original large graph to obtain a sampled graph, and then
estimates the global clustering coefficient by checking the closure of
wedges selected in the sampled graph. The estimate of the total number
of wedges in the original graph is obtained by applying the second
moment estimation method in \cite{ThoZha2012}. 
This algorithm has a
large memory requirement and cannot provide a real-time update of the
estimates. De Stefani {\em et al.} \cite{DBLP:journals/corr/SteEpa2016}
propose a method based on reservoir sampling called 
\algoname{tri\`est}, which also works on fully dynamic graphs. It
adopts random pairing \cite{RaiWol2008}, an extension of the reservoir
sampling, to solve the problem of accounting for edge deletions. This
algorithm uses a fixed sample size and can keep updating the estimates
during the processing of the graph. Since the publication of our preliminary work in \cite{HanSet2017}, a similar approach has been used in  a recently published report \cite{TurTur2017}. However, they limit their focus on estimating the number of triangles on static graphs, and do not consider the dynamic case.

As presented in \cite{DBLP:journals/corr/SteEpa2016},
\algoname{tri\`est} is significantly better than previously known methods
in terms of accuracy, space requirement and the applicability to fully
dynamic graphs. Even though the graph dataset in most real applications has to be stored somewhere, such as on a server accessible by API queries, the framework assumed by \algoname{tri\`est} does not permit queries to the graph dataset. The Edge Sample and Discard (\algoname{esd}), proposed in this paper, assumes a slightly different but more realistic framework allowing  access to the graph dataset information to help substantially improve both the computational/memory costs and the accuracy. 

\section{The Algorithm}
\label{sec:rationale}
The {\em{Edge Sample and Discard}} (\algoname{esd}) algorithm is designed assuming the framework described in Section \ref{sec:framework}. It works on dynamic graphs where both edge deletions and
additions are considered. Additional information, the neighboring nodes of the sampled edges, is queried. It is also generalizable to the case of
directed graphs, but for clarity of presentation in the paper, we will
use undirected graphs in this paper.

\subsection{Preliminaries and Notation}
\label{sec:graphNote}

Let $G_t=(V_t,E_t)$ represent an undirected simple graph, where $t$ is
the time instant and $t \geq 0$.  $V_t$ and $E_t$ are the node set and
the edge set at time $t$, respectively. At the beginning, we have
$V_0=E_0=\emptyset$.  

Consider a stream $\mathcal{S}$ of $((u,v),\beta)$, where $(u,v)$
denotes the edge which is added to or deleted from the graph, and
$\beta \in \{+1,-1\}$. $\beta =+1$ indicates that edge $(u,v)$ is
added to the graph, and $\beta=-1$ indicates that edge $(u,v)$ is
deleted from the graph. For any $t\ge0$, if a new pair $((u,v),\beta)$
arrives at time $t$, we update $G_{t-1}=(V_{t-1},E_{t-1})$ to
$G_t=(V_t,E_t)$ with the corresponding edge addition or deletion.

For simplicity, we drop $t$ from the notation and denote by
$G=(V,E)$ the most recent update of the graph. Let $\Gamma(v)$ denote
the set of neighbors of node $v\in V$, and let $d(v)=|\Gamma(v)|$
denote the node-degree of $v$. A {\em wedge} is a path of length two,
and a {\em triangle} is a closed wedge (a circular path of length
three). Let $T$ denote the set of triangles in $G$ and let $N_T=|T|$.  

The goal of this work is to monitor the edge stream of a graph and
estimate the value of $N_T$ by examining only a small fraction of the
edges and their neighborhoods.  

%

 \subsection{Edge Sample and Discard}
Algorithm \ref{alg:triangle_general} presents the pseudo-code of Edge
Sample and Discard (\algoname{esd}) to estimate the total number of
triangles given a stream of $((u,v),\beta)$. In our algorithm, we use
a global variable $T_{\mathrm{est}}$ to record the real-time estimate
of the total number of triangles in the current graph. We consider both edge addition and deletion operations are included; however, as described in Figure \ref{fig:system_outline} and as in real-life scenarios, the SNS server assumes responsibility for the maintenance and update of the graph datasets while the information related to the dataset can be queried and obtained by \algoname{esd}.

Lines 1-2 in the pseudo-code perform necessary initializations. Lines
3-8 show that for each pair $((u,v),\beta)$ in the stream, we check
the value of $\beta$ and decide whether edge addition or deletion
should be performed. Lines 9-13 perform edge sampling and estimate. We
use the sampling fraction $\alpha$ as the selecting probability. Each
arriving edge has a probability $\alpha$ of being sampled. If an edge
is sampled, the {\em UpdateCount} routine is called. Note that
{\em UpdateCount} works on the graph where the addition or deletion
has just been made. Suppose, at time $t$, an edge $e=(u,v)$ is sampled
and {\em UpdateCount}$(u,v,\beta)$ is called. Then, we examine the
size of the neighborhood of $u$. For $\beta=-1$, when $(u,v)$ is
deleted from $G_{t-1}$, we check whether node $u$ has neighbors in
$G_t$. For $\beta=+1$, where $(u,v)$ is 
added to $G_{t-1}$, we check whether node $u$ has more than one
neighbor in $G_t$. If one of the requirements is fulfilled, we check
the value of $\beta$ and perform the corresponding neighborhood
selection. If $\beta=+1$, we pick one node from the neighbor set of
$u$ other than $v$. For example, we select node $a$ from
$\Gamma(u)\setminus\{v\}$, and thus the probability of $a$ being
picked uniformly at random is $\frac{1}{d(u)-1}$. Since the
probability of sampling edge $(u,v)$ is $\alpha$, the total
probability $P$ of selecting $(u,v)$ and then $a$ as a neighbor of $u$
is:
\begin{eqnarray}
\label{equ:prob}
P=\frac{\alpha}{d(u)-1}. 
\end{eqnarray}

\newcommand{\algrule}[1][.2pt]{\par\vskip.3\baselineskip\hrule height #1\par\vskip.3\baselineskip}
\begin{algorithm}[!t]
\caption{The \algoname{esd} algorithm}\label{alg:triangle_general}
\begin{algorithmic}[1]
\Require{A graph stream $\mathcal{S}$ and sampling fraction $\alpha$}
\State{$T_{\mathrm{est}}\leftarrow 0$}
\State{Create an empty graph $G$}
\For{each pair $((u,v),\beta)$ in $\mathcal{S}$}
\If{$\beta=+1$}
\State{Add new edge $(u,v)$ to graph $G$}
\Else
\State{Delete old edge $(u,v)$ from graph $G$}
\EndIf
\State{$r\leftarrow$ Random number $\in [0,1] $} 
\If{$r\leq \alpha$}
\State{UpdateCount$(u,v,\beta)$}
\State{UpdateCount$(v,u,\beta)$}
\EndIf
\EndFor
\State{\textbf{return} $T_{\mathrm{est}}$}
\end{algorithmic}

\algrule
\begin{algorithmic}[1]
\Statex{Function used in the \algoname{esd} algorithm }
\algrule
\Statex{\textbf{{\em UpdateCount}$(u,v,\beta)$}:}
\If{$|\Gamma(u)|>\frac{1+\beta}{2}$}
\If{$\beta=+1$}
\State{Pick random node $a$ uniformly from $\Gamma(u)\setminus\{v\}$}
\If{$a\in \Gamma(v)$}
\State{$T_{\mathrm{est}}=T_{\mathrm{est}}+\frac{1}{2}\frac{d(u)-1}{\alpha}$}
\EndIf
\Else
\State{Pick random node $a$ uniformly from $\Gamma(u)$}
\If{$a\in \Gamma(v)$}
\State{$T_{\mathrm{est}}=T_{\mathrm{est}}-\frac{1}{2}\frac{d(u)}{\alpha}$}
\EndIf
\EndIf
\EndIf
\end{algorithmic}
\end{algorithm} 

Given a wedge $a$-$u$-$v$, we check whether the closing edge
$(a,v)$ exists by examining $\Gamma(v)$. If $a \in \Gamma(v)$, we
update the triangle estimator $T_{\mathrm{est}}$. The estimate is
updated by applying Eq. (\ref{equ:est_localtriangle}) (in Section
\ref{sec:estvar}). On the other hand, if $\beta=-1$, since edge
$(u,v)$ is already deleted from $G$ and $v$ is no longer a neighbor of
$u$, we pick one node from the neighbor set of $u$. Thus, the total
probability $P$ of selecting $(u,v)$ and then one node from the
neighborhood of $u$ is  
\begin{eqnarray}
\label{equ:prob2}
P=\frac{\alpha}{d(u)}.
\end{eqnarray}

After selecting a node from the neighbor set of $u$, we check whether
the selected node is also a neighbor of $v$. If it is, which means
the subgraph induced by the two incident nodes of the deleted edge
$(u,v)$ and the selected node together is a triangle in $G_{t-1}$, we
update the triangle estimator $T_{\mathrm{est}}$. 

Given a dynamic graph, \algoname{esd} avoids using extra space for storing the sample graph by discarding the processed
edge and nodes after updating the estimate. The total number of
triangles is estimated from the probabilities with which an edge and one
of its neighbor nodes are sampled, and whether the sampled edge and
the node form a triangle. \algoname{esd} can provide a real-time
estimate of the triangle counts in a dynamic graph as new edges come
in or old edges are deleted.  

\section{Quality of Estimation}
\label{sec:estvar}

In this section, we present the mathematical reasoning behind our
triangle estimator and prove that our algorithm provides an unbiased
estimate of the total 
number of triangles with a theoretically tight bound on the
variance. We first discuss the case of dynamic graphs allowing only
the addition of edges and with edge deletions not considered;
we next show that estimating the number of triangles in this
additions-only case is not different from that in the case of a fully
dynamic graph with both additions and deletions.

Let an ordered tuple $(u,v,z)$ denote the sampled edge $(u,v)$ and the
node $z$ selected by {\em UpdateCount}$(u,v,\beta)$. The first element in the
tuple, $u$, is one of the incident nodes of the edge and the second element
is the other incident node.
The
third element of the tuple, $z$, is the node picked from the neighborhood of the first
element, $u$. For example, given a sampled pair $((a,b),+1)$ from the
edge stream, we select node $c$ from $\Gamma(a)\setminus\{b\}$ which
gives us the ordered tuple $(a,b,c)$. 

Let's consider the partially dynamic case with edge additions
only. Suppose we have a stream $\mathcal{S}$ of pairs $((u,v),\beta)$ where $\beta=+1$ for each pair
in $\mathcal{S}$. If a pair $((u,v),+1)$ arrives at time $t$, the
graph $G_{t-1}$ is updated to $G_{t}$ as follows:
\begin{equation*}
G_t=(V_{t-1}\cup\{u,v\},E_{t-1}\cup\{(u,v)\}).
\end{equation*} 
Let $T_t$ denote the set of triangles in $G_t$, where $T_0=\emptyset$. Suppose, at time $t+1$, we get $(e_{t+1},+1)$ from the stream, where $e_{t+1}=(u,v)$ is an edge arriving at time $t+1$ , so we have an updated graph $G_{t+1}$. Let $H(e_t,G_t)$ denote the set of triangles composed of edge $e_t$ in graph $G_t$. We can obtain that $T_{t+1}=T_t\cup H(e_{t+1},G_{t+1})$. Since $T_t\cap H(e_{t+1},G_{t+1})=\emptyset$, 
\begin{eqnarray}
\label{equ:add_set}
|T_{t+1}|=|T_t|+|H(e_{t+1},G_{t+1})|.
 \end{eqnarray}
According to Eq. (\ref{equ:add_set}), we can obtain:
\begin{eqnarray}
\label{equ:total_triangle}
|T_{t}|=\sum_{i=0}^t{|H(e_i,G_i)|}.
 \end{eqnarray}
 
Let $\mathcal{Q}_t$ denote the set of all ordered tuples $(u,v,z)$
that have a non-zero probability to be observed when processing a new
arriving pair $(e_t,\beta)$ from stream $\mathcal{S}$. Let
$\mathcal{T}_t\subseteq \mathcal{Q}_t$ be the set of all ordered
tuples in $\mathcal{Q}_t$ of which the three elements form a triangle
in $G_t$. Note that $(u,v,z)$ and $(v,u,z)$ are two different tuples
but the three elements in each of them induce the same triangle in the graph.  
 
Suppose $(e_t,\beta)$ is sampled, and then one node is picked from the neighborhood of each incident node of $e_t$. So the same triangle may be observed twice during the sampling period. Thus, we can obtain
\begin{equation*}
|\mathcal{T}_t|=2|H(e_{t},G_{t})|.
\end{equation*}

Consider $\mathcal{T}_t' \subseteq \mathcal{T}_t$ as the set of ordered tuples obtained by sampling $(e_t,+1)$ , where the three elements of each ordered tuple in $\mathcal{T}_t' $ form a triangle in $G_t$. Let $P(r)$ be the probability that an ordered tuple $r=(u,v,z) \in \mathcal{T}_t'$ is sampled. By adopting the Horvitz-Thompson construction\cite{HorTho1952}, we come up with the linear estimator,
\begin{equation}
\label{equ:est_localtriangle}
H^t_{\mathrm{est}} = \omega \sum_{r \in \mathcal{T}_t'} \frac{1}{P(r)}.
\end{equation}
$\omega$ is a weight parameter, and $\omega=|H(e_{t},G_{t})|/|\mathcal{T}_t| =1/2$.

Let $\Delta_k=(u,v,z)$ be an element in $\mathcal{T}_t$, where $k \in [1,|\mathcal{T}_t|]$. Remember that the subgraph induced by the three elements of $\Delta_k$ in $G_t$ is a triangle. Let $\delta_k$ denote the existence of
$\Delta_k$ in the set $\mathcal{T}_t'$. We have  
\begin{equation*}
\delta_k=\left\{
\begin{array}{ll}
 1 &    \mbox{ $\Delta_k \in \mathcal{T}_t'$}\\
 0 &   \mbox{ $\Delta_k \notin \mathcal{T}_t'$}
\end{array} 
\right. 
\end{equation*}
Taking the expectation of $H^t_{\mathrm{est}}$, 
\begin{eqnarray*}
\mathrm{E}[H^t_{\mathrm{est}}] &= &\omega \sum_{k=1}^{|\mathcal{T}_t|}\mathrm{E}[\delta_k \cdot \frac{1}{P(\Delta_k)}] \\
&=&\omega \sum_{k=1}^{|\mathcal{T}_t|} P(\Delta_k)\cdot \frac{1}{P(\Delta_k)} \\
  &=&|H(e_{t},G_{t})|
\end{eqnarray*}
The expected number of triangles composed of edge $e_t$ obtained by
our estimator is equal to the actual number of triangles composed of
edge $e_t$ in $G_t$. Applying Eq. (\ref{equ:total_triangle}) and
Eq. (\ref{equ:est_localtriangle}), we get:
\begin{eqnarray*}
|T_t|_{\mathrm{est}}=\omega\sum_{i=0}^t \sum_{r \in \mathcal{T}_t'} \frac{1}{P(r)}
\end{eqnarray*}
According to the linearity of the expectation, $\mathrm{E}[|T_t|_{\mathrm{est}}]=|T_t|$. So we have an unbiased estimator to approximate the total number of triangles in $G_t$.
The variance of $H^t_{\mathrm{est}}$ is:
 \begin{eqnarray*}
 \mathrm{Var}[H^t_{\mathrm{est}}] &= & \mathrm{E}[(H^t_{\mathrm{est}}-\mathrm{E}[H^t_{\mathrm{est}}])^2]\\
 &=& \mathrm{E}[{H^t_{\mathrm{est}}}^2]-|H(e_{t},G_{t})|^2
  \end{eqnarray*}
Expanding $ \mathrm{E}[{H^t_{\mathrm{est}}}^2]$, we get:
 \begin{eqnarray*}
 \mathrm{E}[{H^t_{\mathrm{est}}}^2] &=&\omega^2\,\mathrm{E}\left[\sum_{i=1}^{|\mathcal{T}_t|} \left(\frac{\delta_i}{P(\Delta_i)}\right)^2+\sum_{i\neq j}^{|\mathcal{T}_t|}\frac{\delta_i \delta_j}{P(\Delta_i)P(\Delta_j)}\right] \end{eqnarray*}
Suppose the sampling fraction is $\alpha$, and the maximum degree in
$G_t$ (the graph upon the most recent update) is $d_\mathrm{max}$, then we have $P(\Delta_i) \geq \frac{\alpha}{d_\mathrm{max}-1}$ for any $i \in [1,|\mathcal{T}_t|]$.
Letting $m=\frac{\alpha}{d_\mathrm{max}-1}$, we can obtain, 
 \begin{eqnarray*}
 \mathrm{E}\left[\sum_{i=1}^{|\mathcal{T}_t|} \left(\frac{\delta_i}{P(\Delta_i)}\right)^2\right]\leq|\mathcal{T}_t|\frac{1}{m}
 \end{eqnarray*}
According to the definition of $\mathcal{T}_t$, each ordered-tuple in $\mathcal{T}_t$ represents the triangle which is part of edge $e_t$ in $G_t$. So the two triangles represented by any two ordered-tuples $\Delta_i$ and $\Delta_j$ in set $\mathcal{T}_t$, where $i\neq j $, must have edge $e_t$ as a shared edge. Thus, 
\begin{eqnarray*}
\mathrm{E}\left[\sum_{i\neq j}^{|\mathcal{T}_t|}\frac{\delta_i \delta_j}{P(\Delta_i)P(\Delta_j)}\right]=\frac{2}{\alpha}\left(\frac{|\mathcal{T}_t|}{2}\right)^2
\end{eqnarray*}
So the variance of our estimator is bounded as follows,
\begin{eqnarray*}
\mathrm{Var}[H^t_{\mathrm{est}}] \leq \frac{|\mathcal{T}_t|(d_\mathrm{max}-1)}{4\alpha}+\left(\frac{1}{2\alpha}-1\right)\left(\frac{|\mathcal{T}_t|}{2}\right)^2
\end{eqnarray*}
 Since $\mathrm{Var}[|T_t|_{\mathrm{est}}] =\sum_{i=0}^t \mathrm{Var}[ H^t_{\mathrm{est}}]$, we have
 \begin{eqnarray*}
\mathrm{Var}[|T_t|_{\mathrm{est}}] \leq \frac{|T_t|(d_{\mathrm{max}}-1)}{2\alpha}+\sum_{i=0}^t|H(e_{t},G_{t})|^2 \left(\frac{1}{2\alpha}-1\right)
 \end{eqnarray*} 
 
For any $t>0$, when the degree of the incident nodes of the sampled
edge in $G_t$, where $t$ is the time step that the edge is sampled,
are all equal, the equality in the above bound holds. Therefore, the
bound derived above on the variance of our estimate,
$\mathrm{Var}[|T_t|_{\mathrm{est}}]$, is a strict upper bound.

Further, applying Chebyshev's inequality 
\begin{eqnarray}
 \label{equ:varbound}
P(||T_t|_{\mathrm{est}}-|T_t|| \geqslant \epsilon |T_t|)\leqslant \frac{\mathrm{Var}[|T_t|_{\mathrm{est}}] }{\epsilon^2|T_t|^2}
\end{eqnarray}
The above shows that the relative error of the estimate is influenced
by the sampling fraction and the properties of the graph. The error is
increased as the value of the sampling fraction is 
decreased and the estimate achieves a better accuracy on a graph with more
triangles. Moreover, the local clustering coefficient also affects the
estimate; the algorithm can achieve a better accuracy on a graph where
most of the nodes have a higher clustering coefficient. In general,
\algoname{esd} achieves a better estimate on graphs with a higher
global clustering coefficient.  
  
  
The proof for the fully dynamic case with edge deletions is similar to
the case with additions described above. Suppose we keep performing
edge addition up to time $t$. Thus, at time $t$, we have a graph
$G_t=(V_t,E_t)$ and $T_t$, the set of triangles in $G_t$. Suppose, at
time $t+1$, we get $(e_{t+1},-1)$ from the stream, indicating an edge
deletion, so we have an updated graph $G_{t+1}$. We can easily obtain
that $T_{t+1}=T_t\setminus H(e_{t+1},G_t)$. In other words, we have 
\begin{eqnarray*}
|T_{t+1}|=|T_t|-|H(e_{t+1},G_t)|.
 \end{eqnarray*}
As proved before, we have an unbiased estimator $H^t_{\mathrm{est}} $ for estimating $|H(e_{t},G_t)|$, so in the edge deletion case, we use the same estimator to estimate the decreased number of triangles caused by deleting $e_{t+1}$.

\subsection{Static graphs}
\label{sec:staticgraph}
Although \algoname{esd} is designed for implementation on dynamic
graphs, it can be easily extended to work on static graphs. In
the dynamic case, a triangle can be detected only when the new coming edge
is the closing edge of a wedge already in the graph; however, in
static graphs, each of the three edges of a triangle appears in the
edge stream, so one triangle can be detected every time the new coming
edge is part of this triangle. Thus, for static graphs, the value of
the weight parameter $\omega$ of the estimator is one-third of the one
in the dynamic case.  

Let $\mathcal{S}'$ be the set of ordered tuples which represent the
triangles observed in the sampling period by an edge stream which
delivers random edges from the static graph. Let $P(s)$ be the
probability that a tuple $s\in \mathcal{S}'$ is sampled. By applying
the Horvitz-Thompson construction, we have:
\begin{eqnarray*}
|T|_{\mathrm{est}}=\omega \sum_{s\in \mathcal{S}' } \frac{1}{P(s)}
 \end{eqnarray*}
where the weight parameter is $\omega=1/6$.
 
\section{Performance Analysis}
\label{sec:simulation}
In this section, we perform a comparative analysis of the performance of \algoname{esd} against \algoname{tri\`est} \cite{DBLP:journals/corr/SteEpa2016} and \algoname{doulion}\cite{TsoKan2009}. Both \algoname{tri\`est} and \algoname{doulion}, like \algoname{esd}, can provide a real-time estimate of the total number of triangles by performing edge sampling on an edge stream of a fully dynamic graph. However, \algoname{tri\`est} and \algoname{doulion} assume a traditional streaming graph framework, where a graph can be processed only via a stream of edges. As our comparative analysis will show, the use of additional information by \algoname{esd} already available and kept updated for graph maintenance as per our framework, substantially improves the accuracy.


We use real, simple and undirected graphs from the Network Repository site \cite{NetRep} to create streams of addition-only graphs and fully dynamic graphs. 
Table \ref{tab:graphdata} summarizes some vital properties of these graphs. We also evaluate our algorithm on two real dynamic graphs from \cite{snapnets} and \cite{yahoo}.

\begin{table}[!t]
\begin{center}
\caption{Properties of the graphs used in the experiments. $|E|$ is the number of edges, $|V|$ is the number of nodes, $N_T$ is the number of triangles and $\eta$ is the global clustering coefficient.}
\vspace{0.05in}
\label{tab:graphdata}
\begin{tabular}{>{\hfil}p{80pt}<{\hfil}|>{\hfil}p{50pt}<{\hfil}>{\hfil}p{50pt}<{\hfil}
>{\hfil}p{50pt}<{\hfil}>{\hfil}p{50pt}<{\hfil}}   \hline
\Xhline{1pt}
Graph   &  $|E|$ & $|V|$ & $N_T$  & $\eta$ \\ \Xhline{1pt}
socfb-UCLA & 7.47e+05 & 2.05e+04 & 5.11e+06  & 0.1431  \\ \hline 
socfb-Wisconsin  & 8.35e+05 &2.38e+04 &  4.86e+06    & 0.1201 \\ \hline
com-Amazon &9.26e+05&5.49e+05&6.67e+05 & 0.2052 \\ \hline 
com-DBLP &1.05e+06 &4.26e+05 & 2.22e+06& 0.3064 \\ \hline
web-Stanford &   1.99e+06 & 2.82e+05  & 1.13e+07  &   0.0086  \\ \hline 
web-Google & 4.32e+06& 8.76e+05  & 1.34e+07 &   0.0552    \\ \hline 
\Xhline{1pt}
\end{tabular}
\end{center}
\end{table}


\subsection{Complexity}
\begin{table*}[!t]
\newcommand{\tabincell}[2]{\begin{tabular}{@{}#1@{}}#2\end{tabular}}
\centering
\caption{The complexity of the three algorithms.}
\label{tab:complexity}
\begin{threeparttable}
\begin{tabular}{|c|c|c|c|c|}
\Xhline{1pt}
Algorithm&Time complexity& \tabincell{c}{Server-side\\time complexity}&Space complexity&  \tabincell{c}{Server-side\\space complexity} \\ \hline
\algoname{esd}&$O(p|E|\log d_G)$&$O(p|E|)$&$O(d_G)$& $O(d_G)$\\
\algoname{doulion}\tnote{1}&$O(p|E|+(p|E|)^{\frac{2\omega}{\omega+1}})$&N/A&$O({V_s}^2)$&N/A\\
\algoname{tri\`est}&$O(M d_s \log|E| \log d_s)$&N/A&$O(M)$&N/A\\ \Xhline{1pt}
\end{tabular}
\begin{tablenotes}
\footnotesize
\item[1] The algorithm given in \cite{TsoKan2009} cannot provide a real-time estimate; it only updates the estimate once after processing the entire graph.
\end{tablenotes}
\end{threeparttable}
\end{table*} 

Fast runtime and small space requirement are two
vital goals of a good sampling algorithm.
\subsubsection{Runtime}
Consider the partially dynamic case where edge additions happen $|E|$
times and there are no edge deletions. Suppose the 
neighbors of each node in a graph are stored in a sorted list. This
allows a determination of whether a node is a neighbor of another
specific node in $O(\log d)$ steps where $d$ is the degree of that
specific node.  

In \algoname{tri\`est}, an edge
reservoir is maintained and updated. Suppose the size of the edge
reservoir is $M$. At time $t$ ($t> M$), the probability of updating
the edge reservoir is $M/t$, so the expected number of times that the
edge reservoir is updated is $M+\sum_{t=M+1}^{|E|}\frac{M}{t}\approx M+M\ln|E|.$ 
Each time the edge reservoir is updated,
 \algoname{tri\`est} checks the number of triangles composed of the
 newly sampled edge in the sample graph. Suppose $d_s$ is the maximum
 degree in the sample graph, the computational complexity of
 \algoname{tri\`est} is $O(M d_s \log|E| \log d_s)$. 

As presented in \cite{TsoKan2009}, the computational complexity of
\algoname{doulion} is $O(p|E|+(p|E|)^{\frac{2\omega}{\omega+1}})$,
where $\omega$ is 2.371 and $p$ is the probability of sampling an
edge. 
 
In the case of  \algoname{ESD}, suppose $p|E|$ is the
number of edges which are sampled, and each query of the neighbor nodes takes O(1). Then, for each sampled edge, we
take a maximum of $O(\log d_G)$ steps 
 to sample a neighboring node and determine if the node
and the sampled edge form a triangle. The complexity of
\algoname{esd}, therefore, is $O(p|E|\log d_G)$. Besides, the server needs to take $O(p|E|)$ to respond to the queries.

Table \ref{tab:complexity} summaries the complexity of the three algorithms for the case with no edge deletions. 
 \algoname{esd} is faster
than \algoname{doulion} when $\log d_G < (p|E|)^{0.41}$ (the number of
edges sampled is not too small). In fact, on all real graphs today
and reasonable sample sizes, \algoname{esd} enjoys a lower
computational complexity than \algoname{doulion}. 
For \algoname{esd} and \algoname{tri\`est}, since $\log |E|>\log d_G$ for large
graphs, \algoname{esd} is faster then \algoname{tri\`est} when
sampling the same number of edges ($p|E|=M$). 

Moreover, for each edge deletion, both  \algoname{doulion} and \algoname{tri\`est} have to check whether the deleted edge is in the sample set or not, and update the estimates. While in \algoname{esd}, it avoids looking up the sample set and processes edge deletions with a sampling probability.
So our algorithm is substantially faster than \algoname{doulion} and \algoname{tri\`est}, especially when facing a large amount of edge deletions.  

Note that  \algoname{esd} uses a different framework from \algoname{doulion} and \algoname{tri\`est}, neither of which involve querying the server for additional information. Given the framework used by \algoname{esd}, it involves an additional server-side cost in responding to the queries. As shown in Table \ref{tab:complexity}, \algoname{esd} achieves a substantially better trade-off saving computational and memory costs with a small amount of extra effort on the part of the server.

\subsubsection{Space}
At first sight, it may appear as though the framework used in this paper to develop the ESD algorithm requires the storage of the entire graph while \algoname{doulion} and \algoname{tri\`est} only have to store the sampled graph. However, this mischaracterizes  the actual storage needs under these frameworks. In real contexts, even streaming graph data for a fully dynamic graph are ultimately generated by a system/server which maintains and keeps updated a graph dataset. After all, a well-maintained graph dataset is essential for the normal functioning of most applications relying on the graph. Before, after and in the midst of any graph analytics, in most real contexts, the graph datasets are still kept stored somewhere. So even for the traditional streaming graph model (used by \algoname{tri\`est} and \algoname{doulion}), the existence and the storage needs of the complete graph dataset {\em cannot} and {\em should not} be ignored. It is unrealistic to assume that, after performing a streaming graph algorithm, one would only store the sampled graph and not store anywhere the large dynamic graph observed so far. Therefore, the cost of storing the dynamic graph is a necessity for all of the three algorithms and their respective frameworks. 

For the graph analytics part, our algorithm avoids the requirement of extra memory to store the sampled edges by performing independent collections of edges and discarding edges after updating the estimators. The list of the neighbor nodes of the sampled edge are required for estimating which leads to a memory requirement $O(d_G)$ where $d_G$ is the maximum degree in the original graph.
As described in Section \ref{sec:framework}, the graph datasets are maintained regardless of whether the graph analytics is applied or not. Thus, for the server-side space complexity, we do not  include the cost for maintaining the graph datasets, and only count the extra space complexity required by our algorithm. 

\algoname{doulion} uses a certain probability $p$ to sample edges in
the stream, and the samples are maintained in the memory during
the entire process. So the amount of memory used is not fixed and partially depends on the algorithm used to calculate the exact number of triangles in the sampled graph. In \cite{TsoKan2009}, the fast matrix multiplication is used to count triangles in the sampled graph, so the space complexity is $O({V_s}^2)$, where $V_s$ is the number of nodes in the sampled graph. \algoname{tri\`est} is a reservoir sampling
based algorithm which uses a fixed amount of memory to store the
sampled edges.

\subsection{Partially dynamic case}
We show the comparison of the performances of \algoname{esd} with \algoname{tri\`est} and \algoname{doulion} on dynamic graphs where only edge additions are considered. The edge stream is generated by permuting the edges uniformly at random. 

We consider the relative error in estimating the triangle number as a measure of the accuracy. The relative error is measured as:
\begin{eqnarray*}
\mathrm{Relative~error} = \frac{\mathrm{Average~estimate} - \mathrm{Actual~value}}{\mathrm{Actual~value}},
 \end{eqnarray*}
where the average estimate is the mean of the estimated value over 100 independent runs. 

\begin{table}[!t]
\centering
\caption{The relative errors in the estimates of the total number of triangles. Sample size is the number of edges sampled.}
\vspace{0.05in}
\label{tab:rel_error}
\begin{tabular}{cc|ccc} 
\Xhline{1pt}
 & &\multicolumn{3}{c}{Triangles $N_T$}   \\ 
Graph   &Sample &\multicolumn{3}{c}{Relative error (\%)}\\ 
 \cline{3-5}
name & size&  \algoname{esd} &\algoname{doulion} & \algoname{tri\`est}  \\ \hline
socfb-UCLA& 7,476&0.0958&1.6274&3.5433\\ \hline
socfb-Wisconsin&8,359& 0.5380&4.2736&2.6817\\ \hline
com-Amazon& 9,258&0.2329&6.4262&3.4856  \\ \hline
com-DBLP&10,498&0.4301&1.5458 &3.4481\\ \hline
web-Stanford  & 19,926&0.9796&3.1822& 5.6751 \\ \hline
web-Google & 43,220 & 0.0336&2.2997&3.2821\\ \hline
\Xhline{1pt}
\end{tabular}
\end{table}

\begin{figure*}[!t]
\centering
    \subfigure[{socfb-UCLA}]{ 
       \includegraphics[width=2.2in]{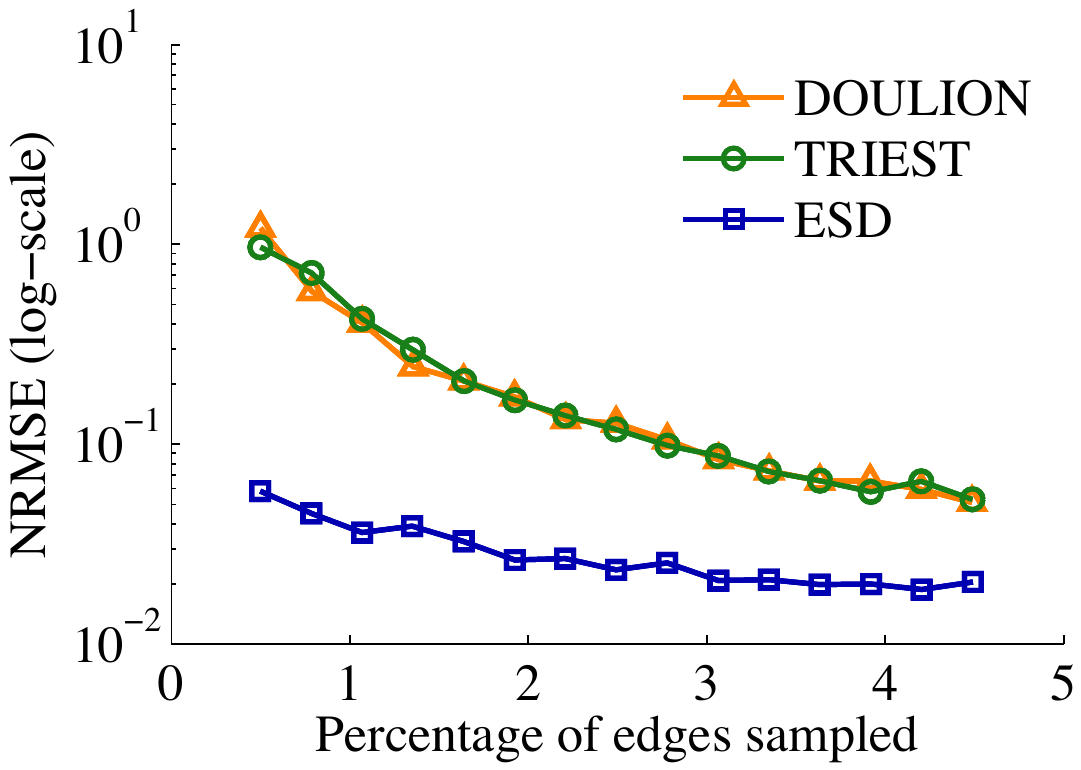}
       }
    \subfigure[{socfb-Wisconsin}]{
       \includegraphics[width=2.2in]{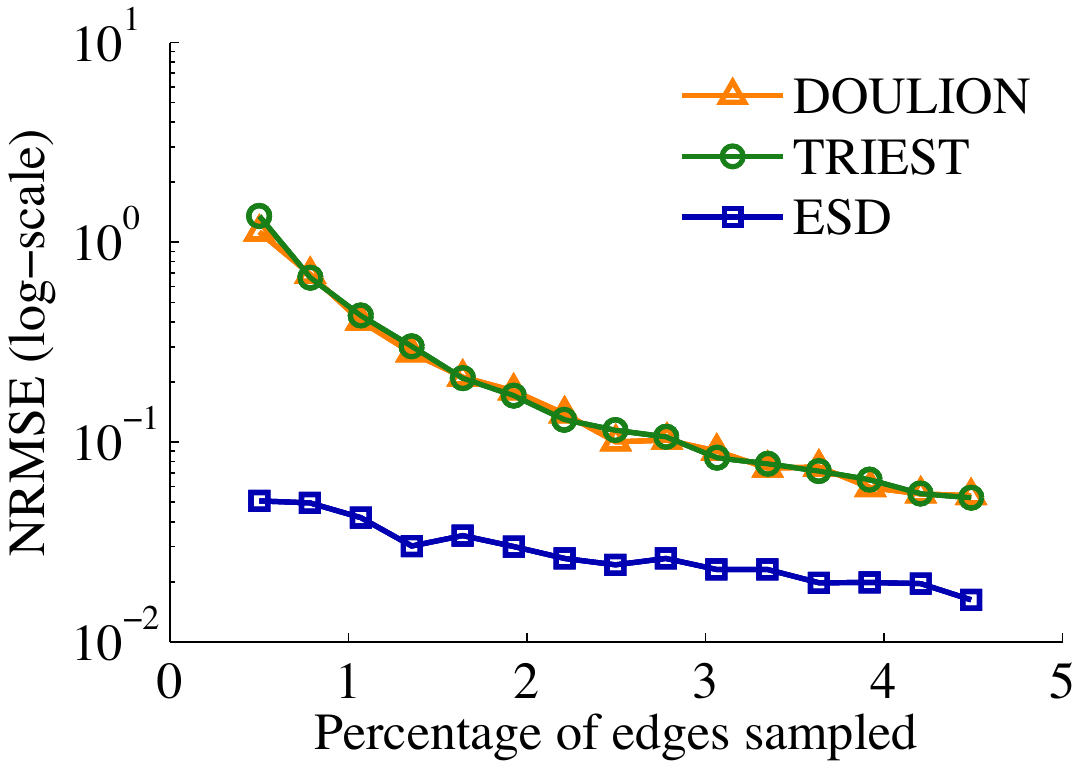}
       }
      \subfigure[{com-Amazon}]{
       \includegraphics[width=2.2in]{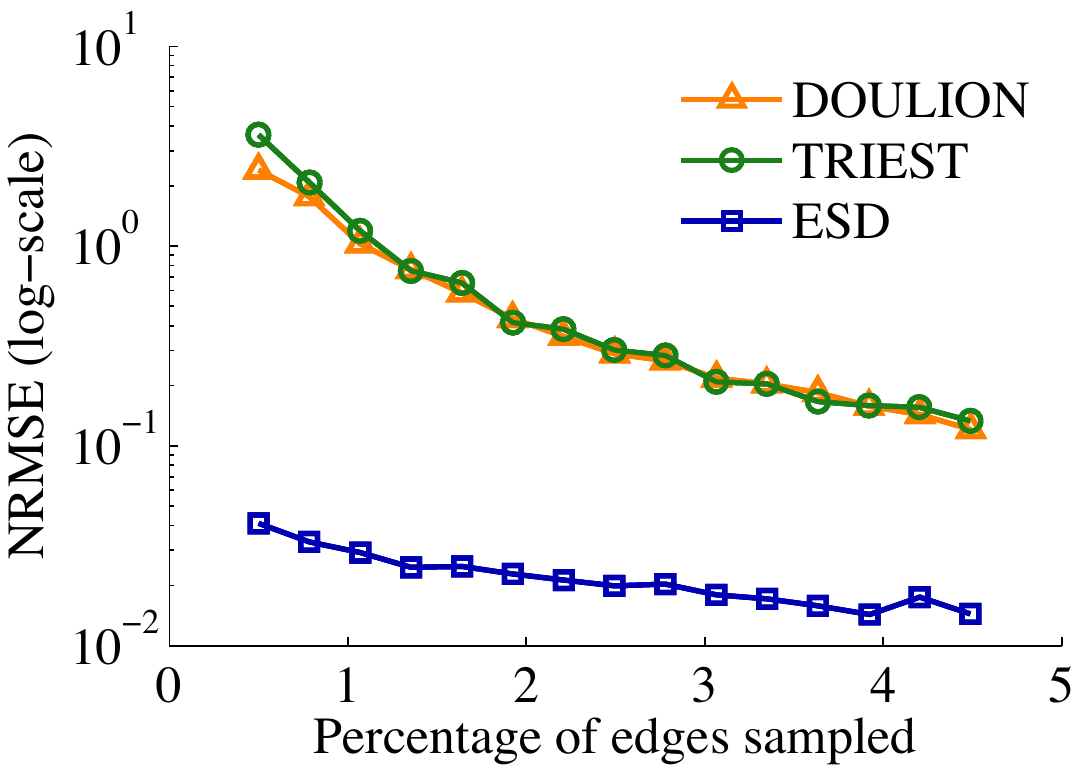}
             }
      \subfigure[{com-DBLP}]{ 
       \includegraphics[width=2.2in]{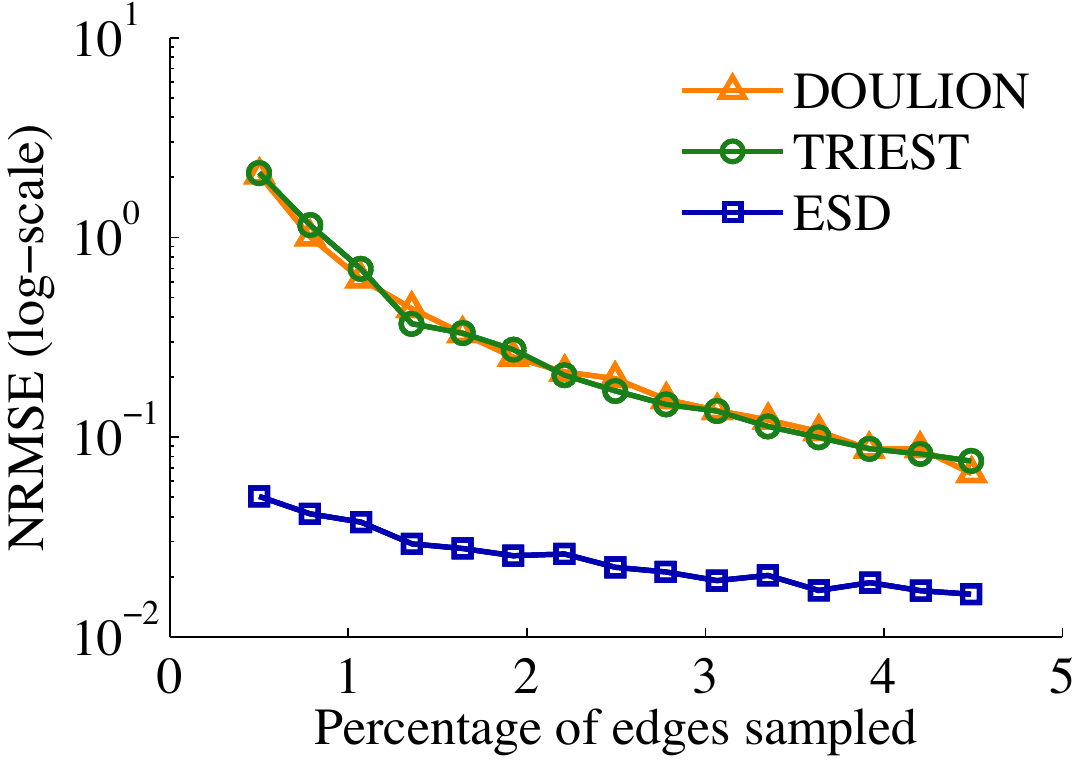}
       }
       \subfigure[{web-Stanford}]{ 
      \includegraphics[width=2.2in]{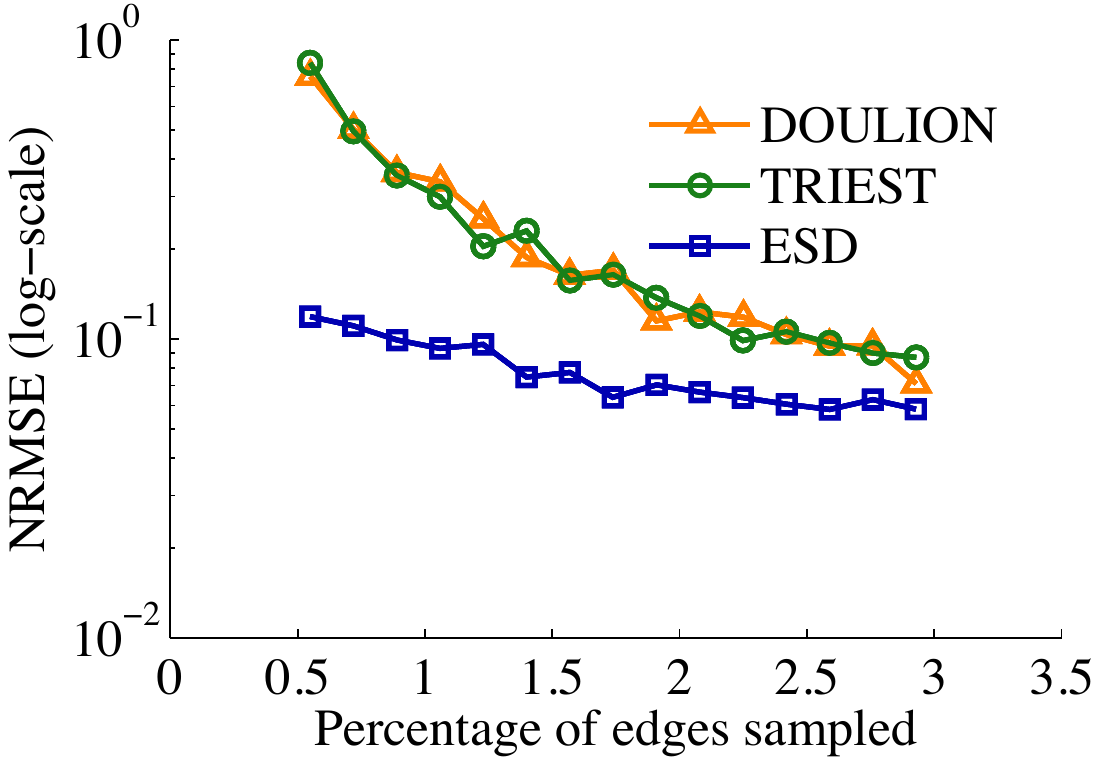}
        }
       \subfigure[{web-Google}]{ 
       \includegraphics[width=2.2in]{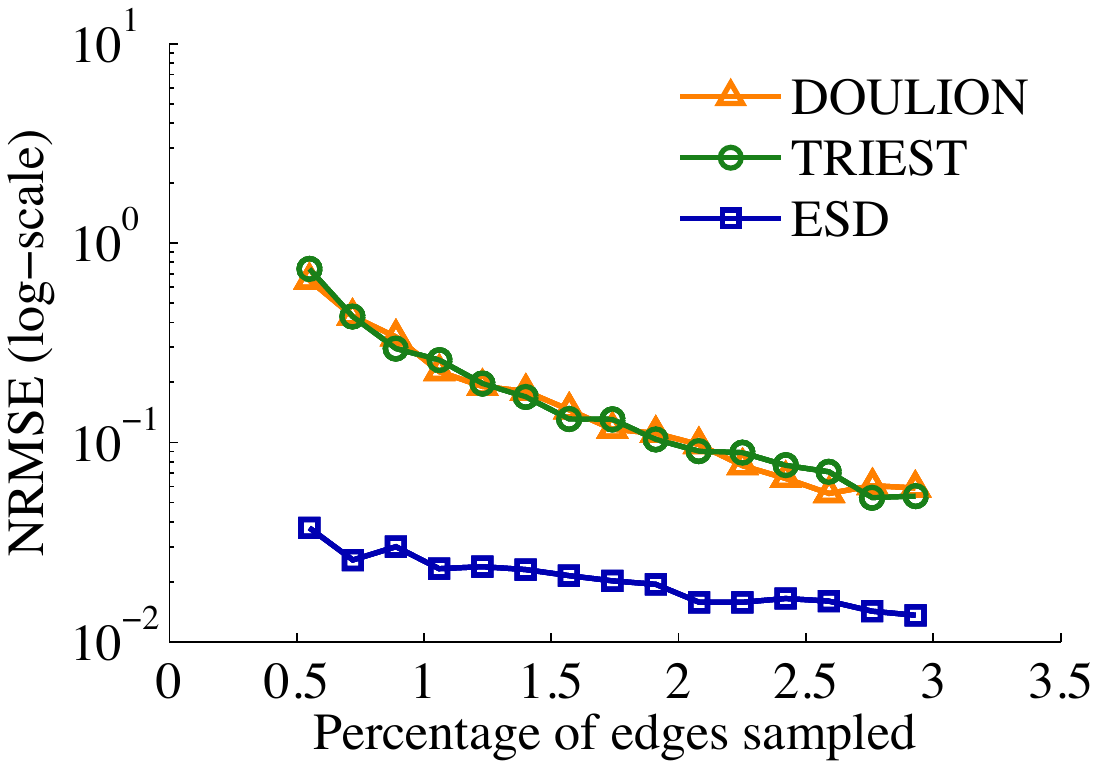}
       }

   \caption{Comparison of NRMSEs of the estimates for three algorithms over 100
     independent runs. }\label{fig:accuracyerrbar} 
\end{figure*} 

Table \ref{tab:rel_error} shows relative errors in estimating the
total number of triangles for each of the three algorithms. We sample
1\% of the edges for each graph. As shown in the table, \algoname{esd}
achieves better accuracy than the other algorithms on all the
graphs. On most of the graphs, the relative errors obtained by \algoname{esd} are at least
one order of magnitude smaller than the errors obtained by
\algoname{doulion} and \algoname{tri\`est}.

To further compare the accuracy of the three algorithms, we use the normalized root mean square error:
\begin{eqnarray*}
\mathrm{NRMSE}=\frac{\sqrt{\mathrm{E}[(\mathrm{estimate}-\mathrm{Actual~value})^2]}}{\mathrm{Actual~value}},
\end{eqnarray*}
Figure \ref{fig:accuracyerrbar} depicts the average NRMSEs based on
100 independent runs for each graph as the sample sizes are
increased. We can see from the figure that \algoname{esd} has the
smallest NRMSEs in all cases. Especially when the sample size is
small, the NRMSEs of \algoname{esd} are almost one order of magnitude
smaller than the NRMSEs of \algoname{doulion} and
\algoname{tri\`est}. On most of the graphs, our algorithm achieves similar NRMSEs to the other two algorithms with one-ninth of the sample size used by these two algorithms. In other words, to achieve equivalent accuracy, \algoname{esd} requires fewer samples, and thus reduces the computational cost.

\subsection{Fully dynamic graphs}
In the experiments for the fully dynamic case, we use the model
presented in \cite{DBLP:journals/corr/SteEpa2016} to simulate the deletions or additions of nodes or edges.

We first tested the performances of the three algorithms on dynamic graphs where both edge additions and deletions are considered. For each test on
the graph, we first generate a stream of edges by randomly permuting
the edges. Initially, an empty graph $G$ is created. The arrival of
each edge in the stream is treated as an edge addition, and each new
edge is added into $G$. A probability $p_e=0.0001$ is used to
decide whether a deletion event should be performed after each edge
addition is made. If a deletion event happens, every edge in $G$ has a
probability $p_d=0.01$ of being deleted.

\begin{figure*}[!t]
\centering
    \subfigure[{socfb-UCLA}]{ 
       \includegraphics[width=2.2in]{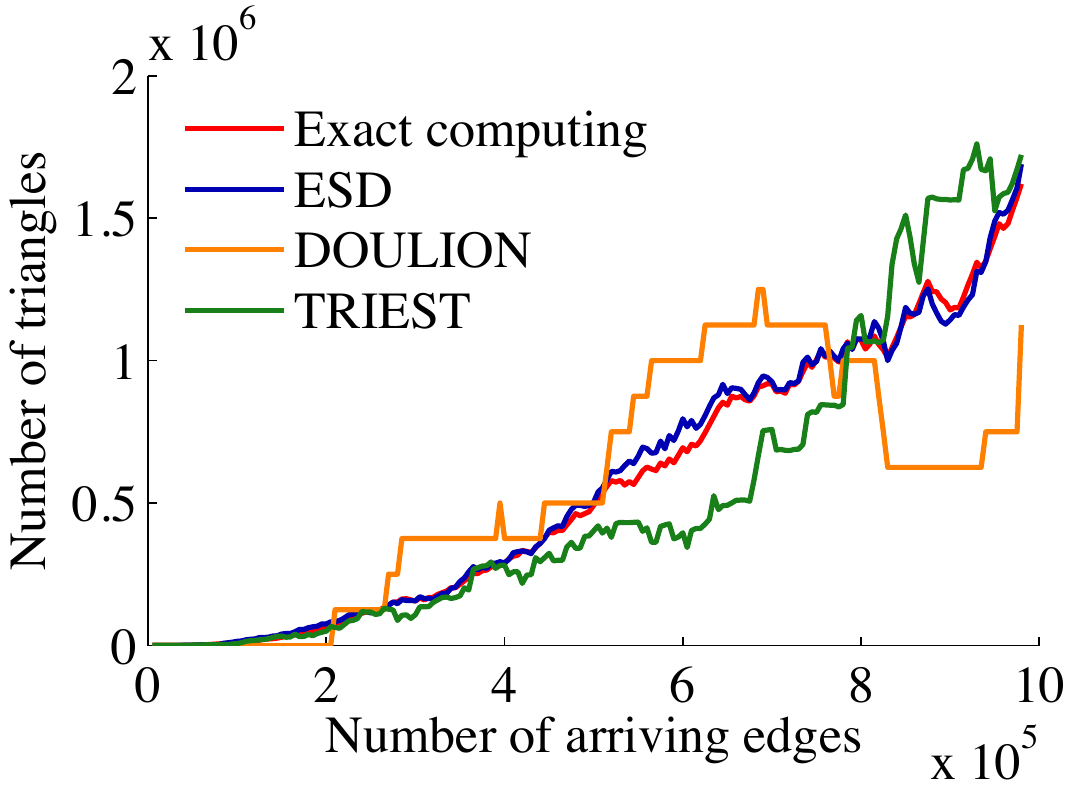}
       }~\subfigure[{socfb-Wisconsin87 }]{
       \includegraphics[width=2.2in]{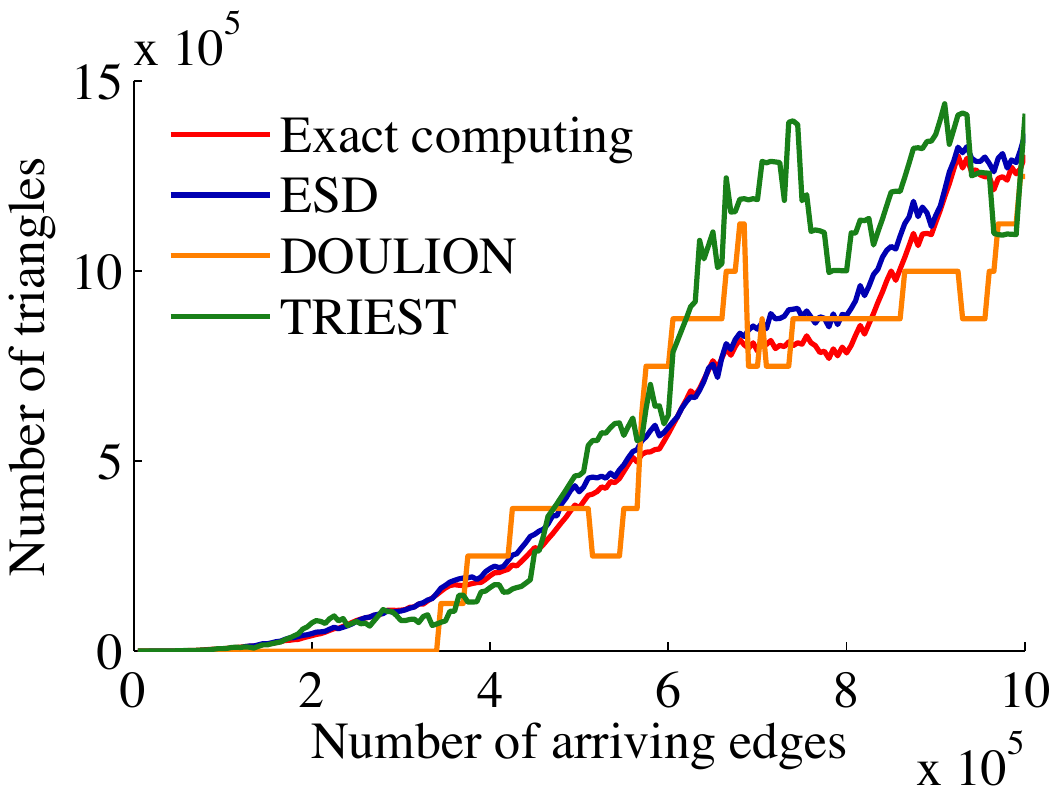}
	} \subfigure[{com-DBLP}]{ 
       \includegraphics[width=2.2in]{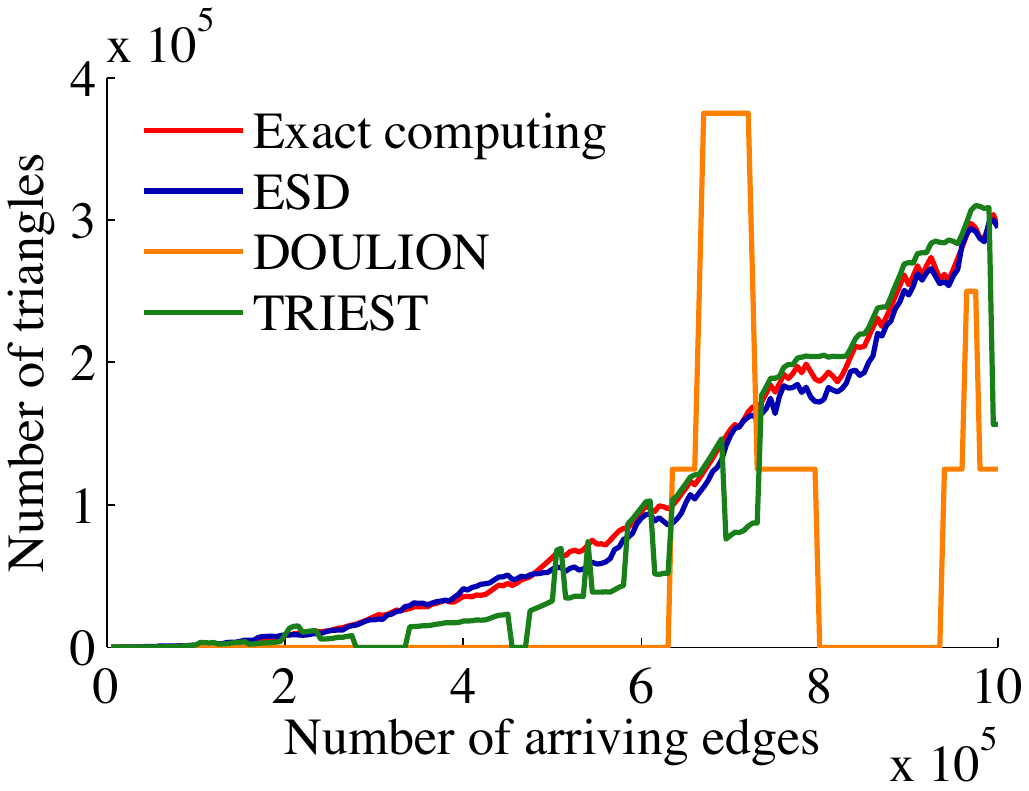}
    }
   \caption{Comparison of the estimated values of
     total number of triangles in dynamic case. }\label{fig:dynamic_test} 
\end{figure*} 

Figure \ref{fig:dynamic_test} shows the comparison of the estimates of
the triangle number for three algorithms. The red line indicates the actual number of triangles
obtained by the exact triangle computing algorithm. For all graphs,
the final sample sizes of both \algoname{doulion} and \algoname{esd}
are approximately equal to 10,000, while the sample size of
\algoname{tri\`est} is fixed at 10,000. 

As shown in the figure, \algoname{esd} has the best performance among
the three in tracking the changes in the number of triangles, even at small sample sizes. In the case of
\algoname{doulion}, however, the edge deletion affects the accuracy of
its estimate. If many deleted edges are edges held in the sample set,
the sample set shrinks quickly, significantly reducing the accuracy of
estimates made by \algoname{doulion}. For example, in the com-DBLP
graph, the estimate obtained by \algoname{doulion} is sometimes more
than twice as large as the actual value. 

\algoname{tri\`est} uses reservoir sampling with a fixed size of the
sample set. If the deleted edge is an edge in the sample set, it would
be removed from the sample set, and the edge deletion in the sample
set would be compensated by future edge insertion. So
\algoname{tri\`est} can maintain a sample set with fixed number of
edges during the entire sampling period. In other words, the number of
edges sampled by \algoname{tri\`est} is always larger than the number
of edges sampled by \algoname{esd} and \algoname{doulion} in the
simulations. The figure shows that \algoname{esd} achieves a better
performance than \algoname{tri\`est} in terms of the accuracy even
though \algoname{tri\`est} samples more edges than \algoname{esd}. 

Besides edge deletion, in the real world, node deletion also occurs. Deletion of a single node can be modeled as a sequence of deletions of edges adjacent to that node. We also tested the performances of the three algorithms on dynamic graphs with node deletions. For each test on the graph, we use a probability $p_e=0.0001$ to decide whether a deletion event should be performed after the occurrence of each edge addition. If a deletion event happens, every node in the current graph has a probability $p_d=0.001$ of being deleted.

\begin{figure*}[!t]
\centering
    \subfigure[{socfb-UCLA}]{ 
       \includegraphics[width=2.2in]{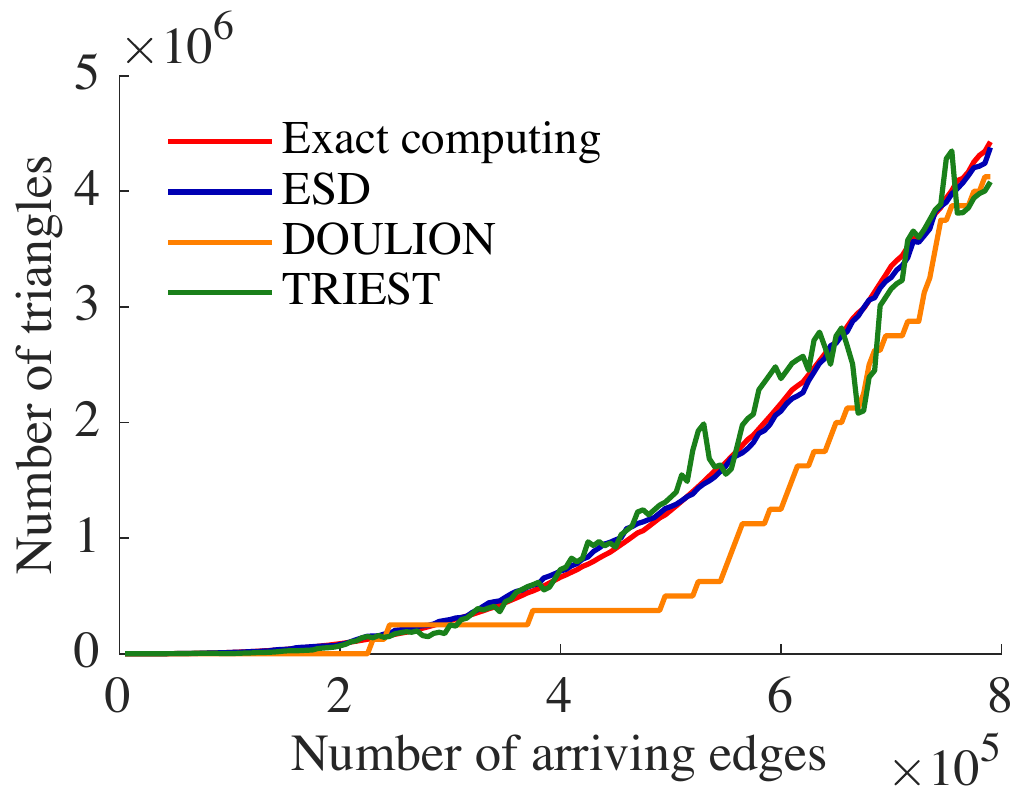}
       }~\subfigure[{socfb-Wisconsin87 }]{
       \includegraphics[width=2.2in]{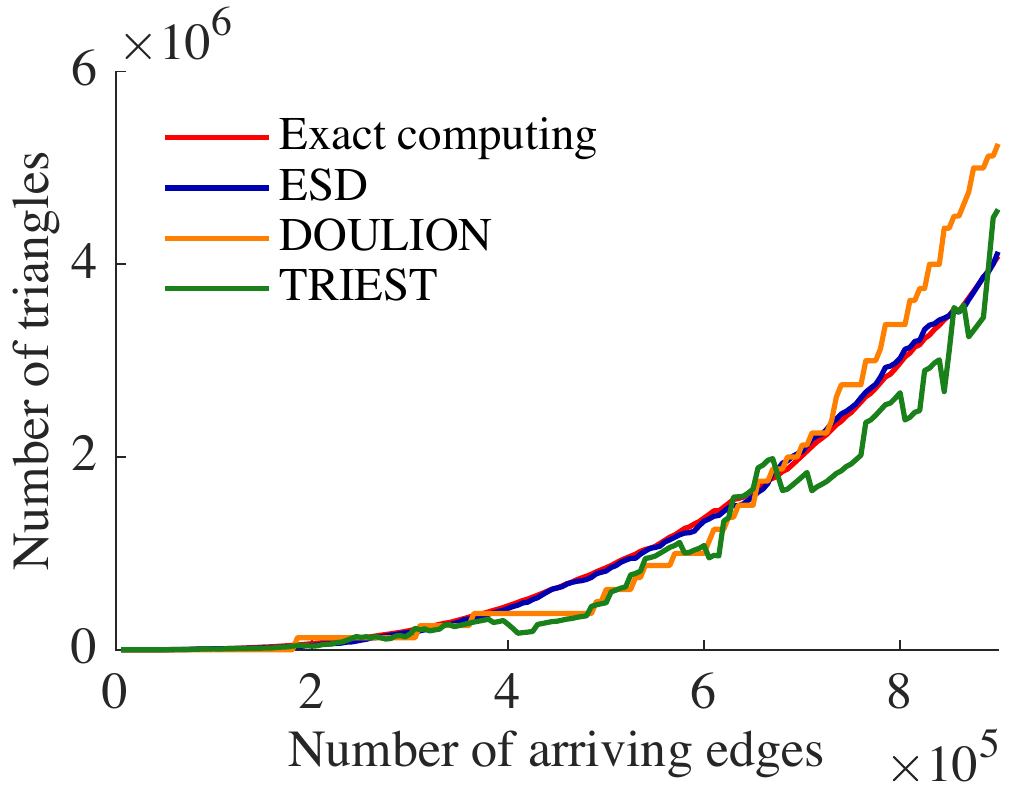}
	} \subfigure[{com-DBLP}]{ 
       \includegraphics[width=2.2in]{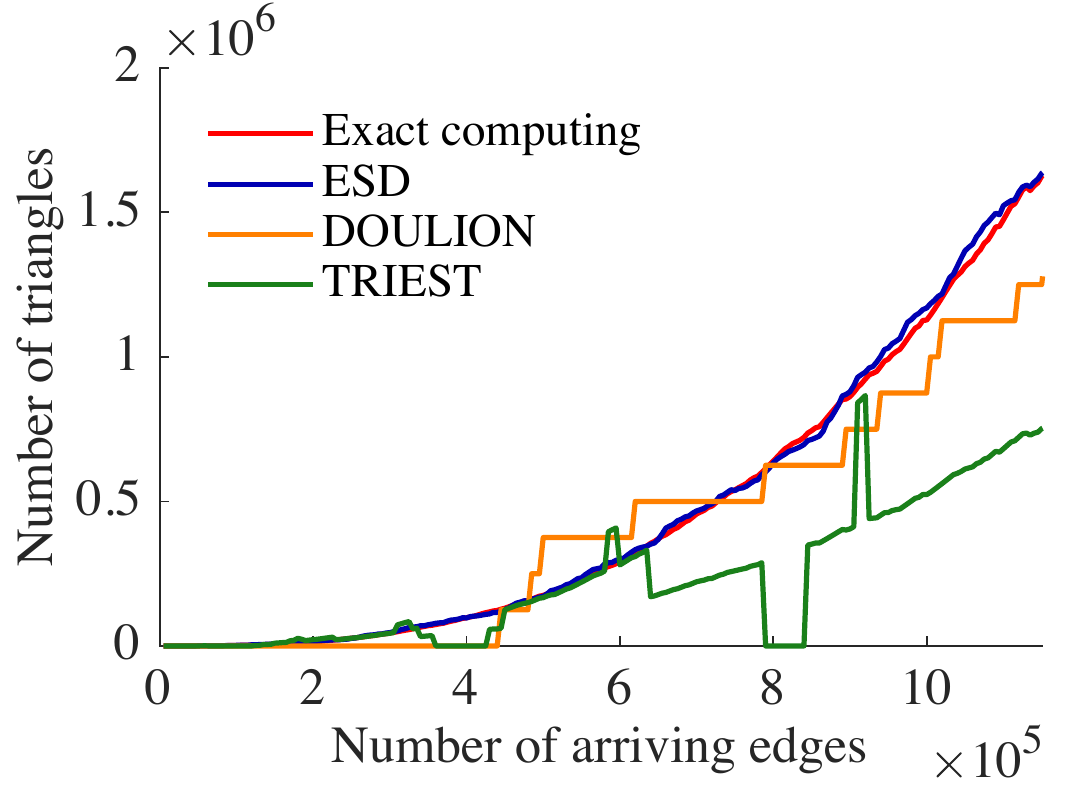}
    }
   \caption{Comparison of the estimated values of
     total number of triangles in dynamic case. }\label{fig:dynamic_test2} 
\end{figure*}

Figure \ref{fig:dynamic_test2} plots the comparison of the estimates of
the number of triangles for the three algorithms. For all graphs,
the sampling probability is set to 0.02 for \algoname{doulion} and is set to 0.01 for \algoname{esd}, while the sample size of \algoname{tri\`est} is fixed at 10,000. 

As plotted in the figure, \algoname{esd} achieves the best performance among
the three in estimating the number of triangles in dynamic graphs with node deletions. The red line which indicates the actual number of triangles, is extremely close to the blue line which plots the estimates obtained by \algoname{esd}. The closeness between the red line and the blue line indicates that our algorithm is capable of providing accurate estimates of number of triangles in a real-time fashion. Moreover, for all graphs, \algoname{esd} has the smallest final sample size. In other words, our algorithm samples fewer edges, but achieves higher accuracy than the other two algorithms in tracking the number of triangles.

Besides the above mentioned models, we also tested the performance of our algorithm on two real dynamic graphs. Oregon-2 dataset \cite{snapnets} contains 9 autonomous system (AS) graphs which represent AS peering information inferred from Oregon route-views. It was collected from March 31, 2001, to May 26, 2001. 
Yahoo! Message dataset \cite{yahoo} which has 28 graphs, was generated by a small subset of Yahoo! Messenger users from different zip codes for 28 days starting from April 1, 2008. Graphs in each of the datasets are timestamped. Each of the graphs is processed sequentially according to its time order. Edges which are not present in the previous graph, but are in the current graph, are treated as edge additions, and edges which are present in the previous graph, but are not in the current graph, are treated as edge deletions. Thus, both of the two datasets exhibit the addition and deletion of the edges over time. 

\begin{figure}[!t]
\centering
    \subfigure[{Oregon-2}]{ 
       \includegraphics[width=3.5in]{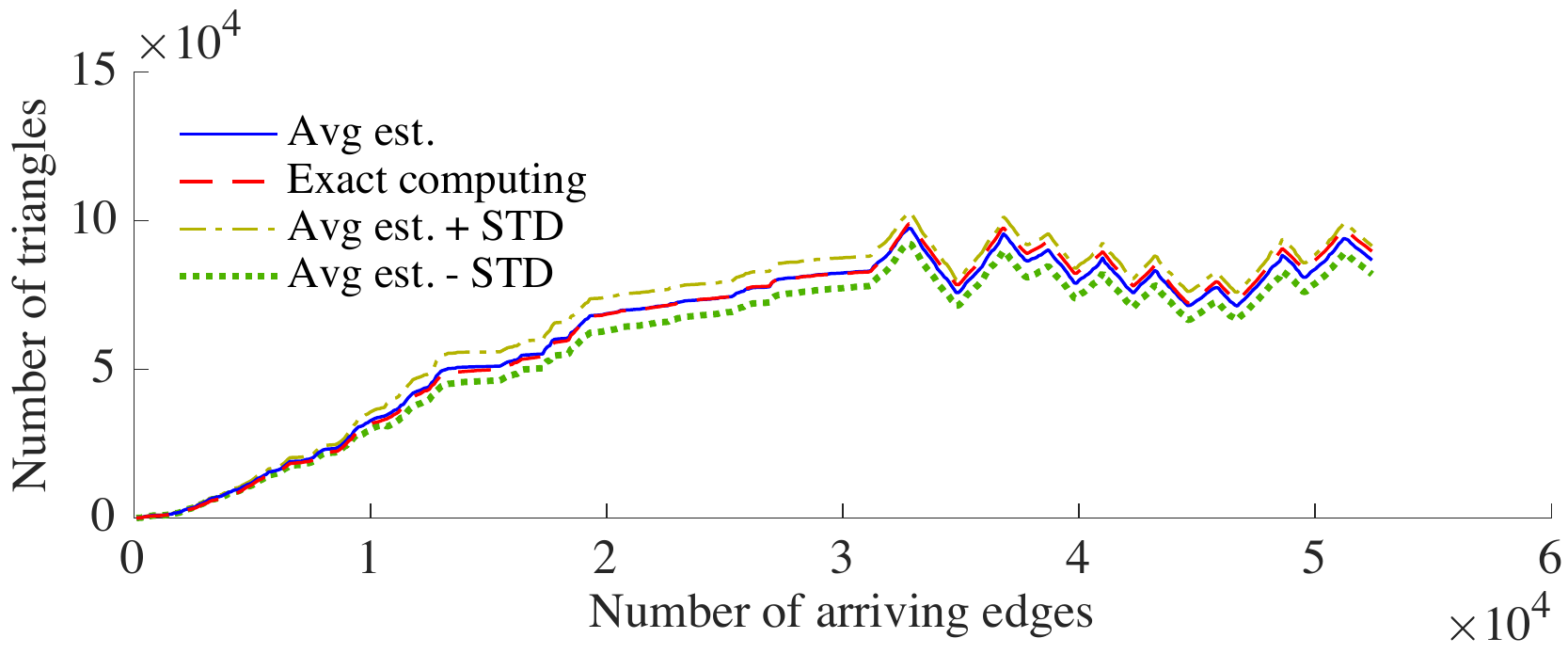}
       }
       \subfigure[{Yahoo! Messenger }]{
       \includegraphics[width=3.5in]{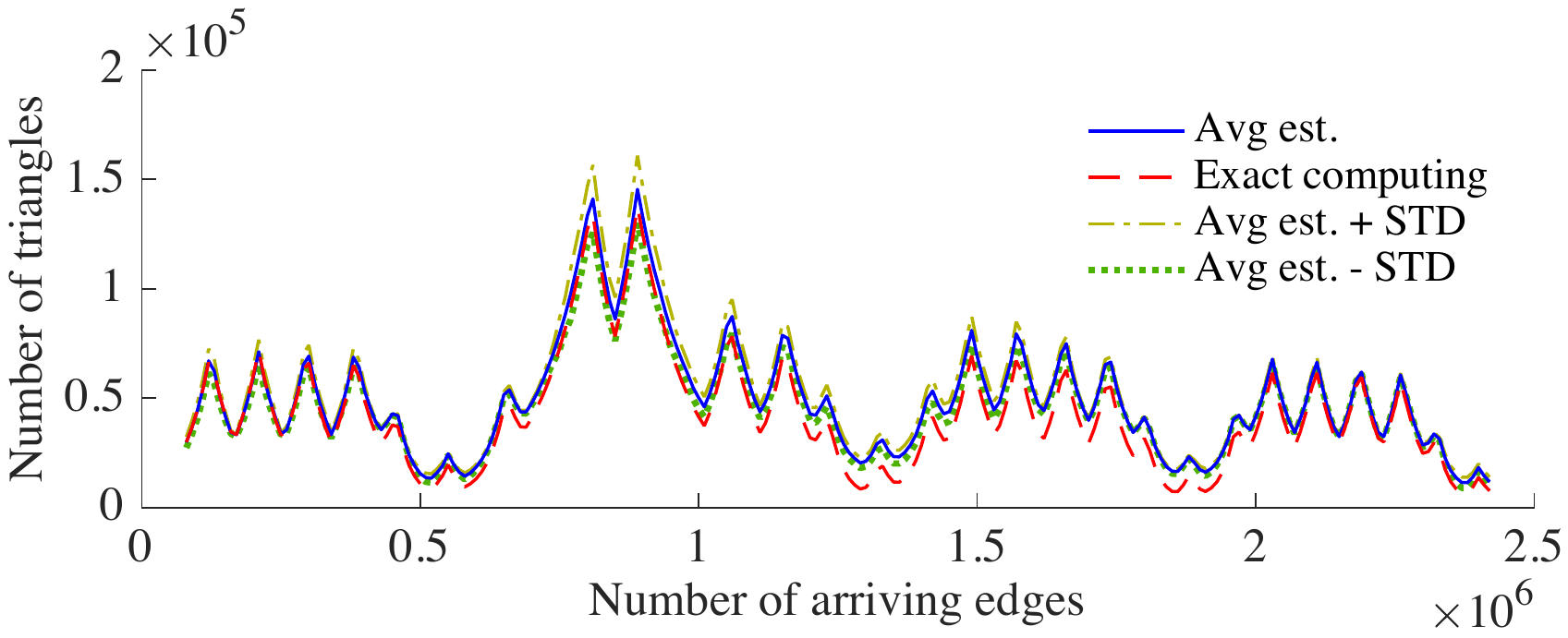}
    }
   \caption{Estimation of the
     total number of triangles in dynamic case. }\label{fig:dynamic_testreal} 
\end{figure}

Figure \ref{fig:dynamic_testreal} shows the estimation of the total number of triangles on real dynamic graphs when the sampling fraction is $\alpha = 0.01$. As shown in the figure,  \algoname{esd} has a good performance on real graphs in terms of the accuracy and the variance. Even on the Yahoo! Message graph, where the number of triangles changes frequently and dramatically, our algorithm is still capable of tracking variation on the number of triangles accurately.

\begin{table}[!h]
\begin{center}
\caption{Properties of the Barab$\mathrm{\acute a}$si-Albert (BA) graphs used in the experiments. $\gamma$ is the power of the preferential attachment, $N_T$ is the number of triangles, and $\eta$ is the global clustering coefficient.}
\vspace{0.05in}
\label{tab:BAdata}
\begin{tabular}{>{\hfil}p{50pt}<{\hfil}|>{\hfil}p{30pt}<{\hfil}>{\hfil}p{50pt}<{\hfil}
>{\hfil}p{50pt}<{\hfil}>{\hfil}p{50pt}<{\hfil}>{\hfil}p{40pt}<{\hfil}}    \hline
\Xhline{1pt}
Graph& $\gamma$ & $N_T$ & $\eta$&$|E|$&$|V|$ \\ \Xhline{1pt}
BA-1  &1.5& 281,296 & 0.00255&200,500& 20,000  \\ \hline 
BA-2  &1.5&  1,046,132   &  0.00381 &399,500 & 20,000 \\ \hline
BA-3 &1.5& 6,874,145&  0.01195 &996,500& 20,000\\ \hline 
BA-4 &1.0& 3,526,361 & 0.02615&1,474,100& 20,000\\ \hline
BA-5   &1.5& 3,473,097    &   0.00818&757,700& 20,000 \\ \hline 
BA-6   &2.0&  3,464,420  &  0.00233 &598,500& 20,000  \\ \hline 
\Xhline{1pt}
\end{tabular}
\end{center}
\end{table}

\subsection{Relationship to graph properties}
We show the influence of the properties of the graph on the
performance of \algoname{esd} by testing on several
Barab$\mathrm{\acute a}$si-Albert (BA) graphs. For ease in
illustrating this relationship, we only consider edge-additions in
this set of experiments. We ran \algoname{esd} on BA graphs
with the same number of nodes, but with different numbers of edges and
different powers of the preferential attachment. For each BA graph, we
start with an Erd$\mathrm{\ddot{o}}$s-R$\mathrm{\acute e}$nzi graph
with 100 nodes. Then, in each time step, one node is added to the
graph, and the new node initiates dozens of edges to old nodes. The
probability that an old node is selected is given by: 
\begin{eqnarray*}
P(i) \thicksim {d_i}^\gamma
\end{eqnarray*}
where $d_i$ is the degree of node $i$ in the current time step and
$\gamma$ is the power of the preferential attachment. Table
\ref{tab:BAdata} lists some basic properties of the BA graphs used in
these simulation experiments.

\begin{figure*}[!b]
\centering
    \subfigure[{BA-1 ($\gamma=1.5$, $N_T=281,296$)}]{ 
       \includegraphics[width=2.2in]{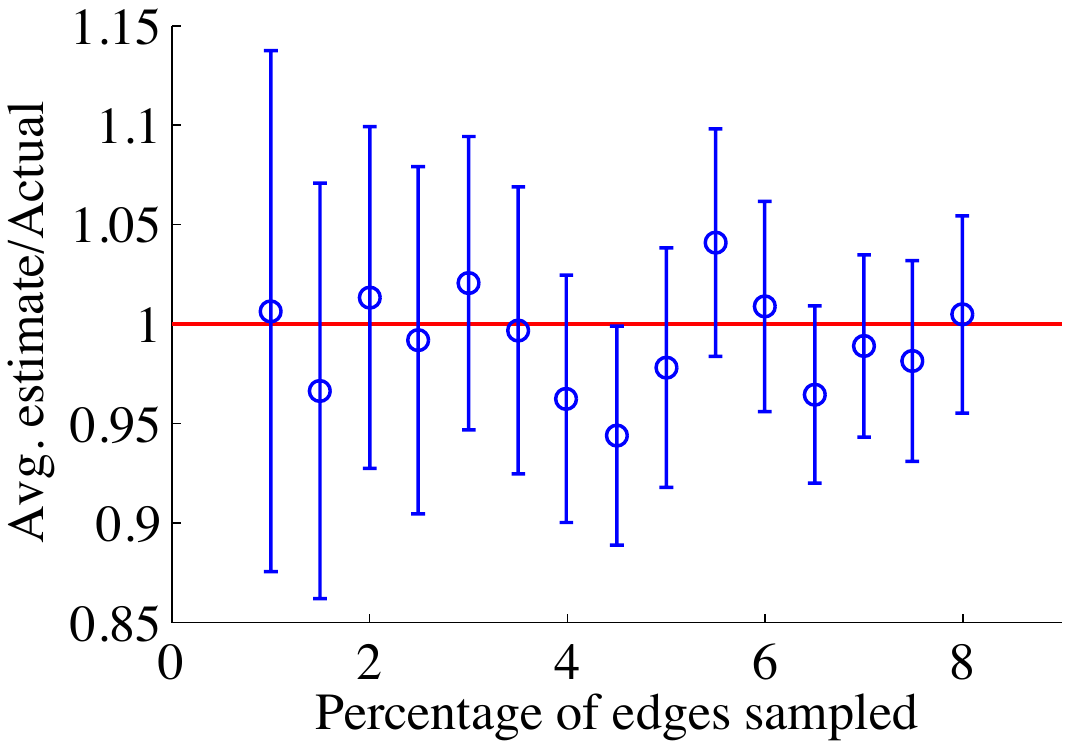}
       }
       \subfigure[{BA-2 ($\gamma=1.5$, $N_T=1,046,132$)}]{
       \includegraphics[width=2.2in]{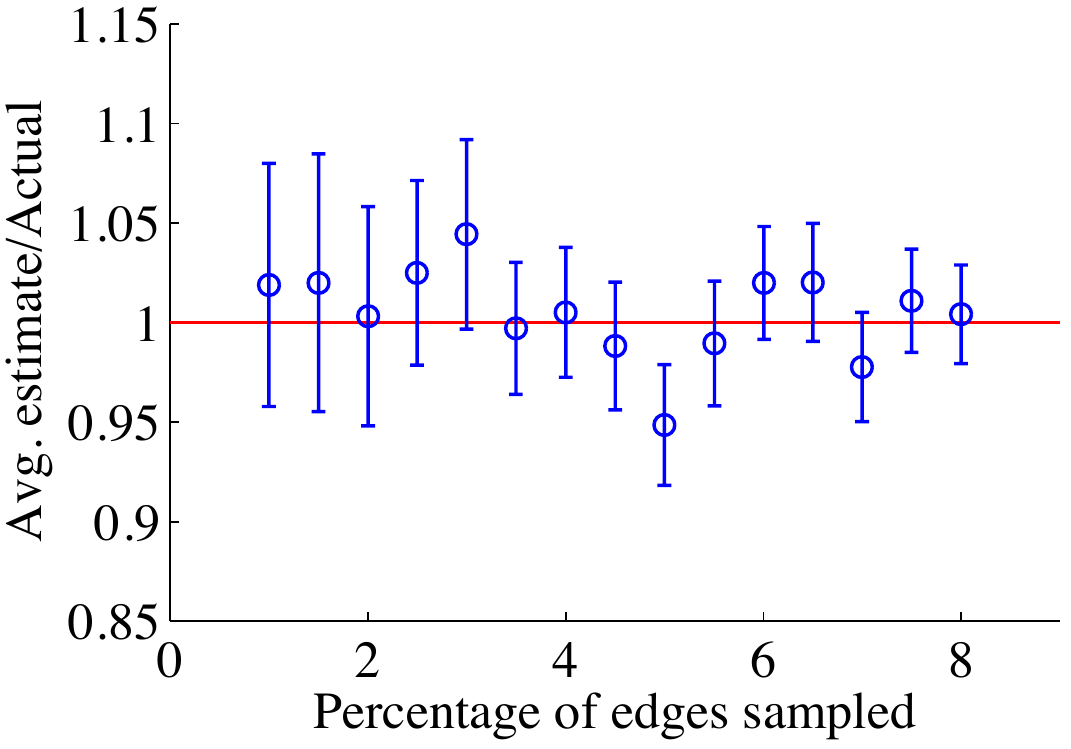}
	}
	\subfigure[{BA-3 ($\gamma=1.5$, $N_T=6,874,145$)}]{ 
       \includegraphics[width=2.2in]{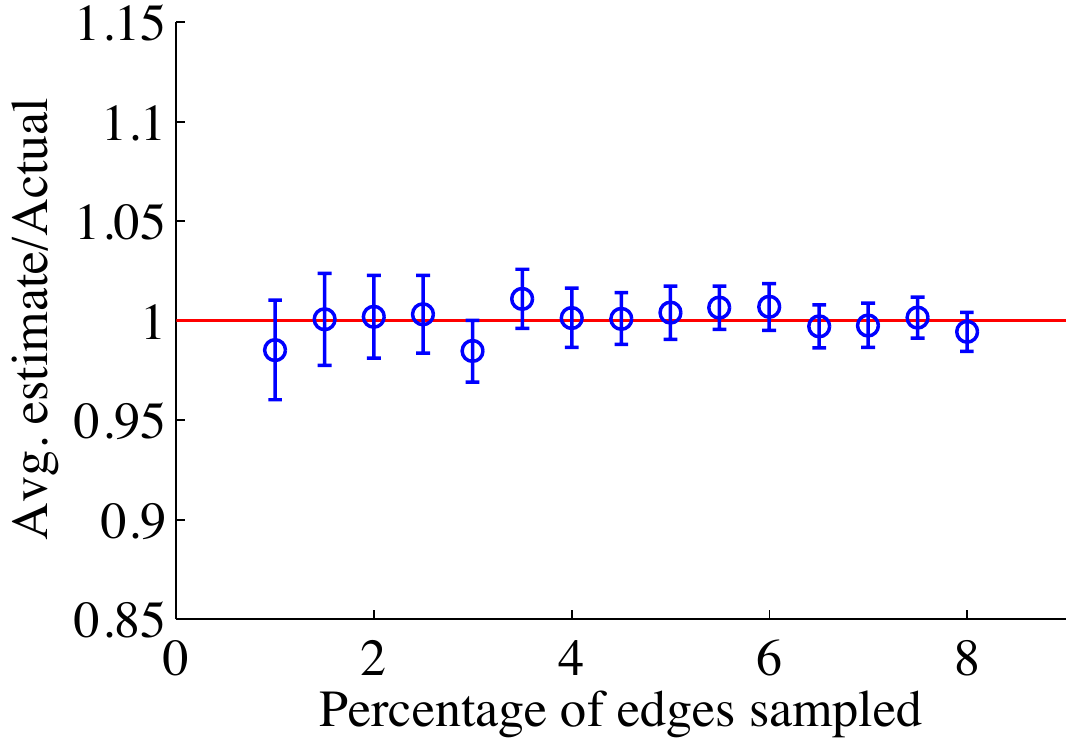}
         }
     \subfigure[{BA-4 ($\gamma=1.0$, $N_T=3,526,361$)}]{ 
       \includegraphics[width=2.2in]{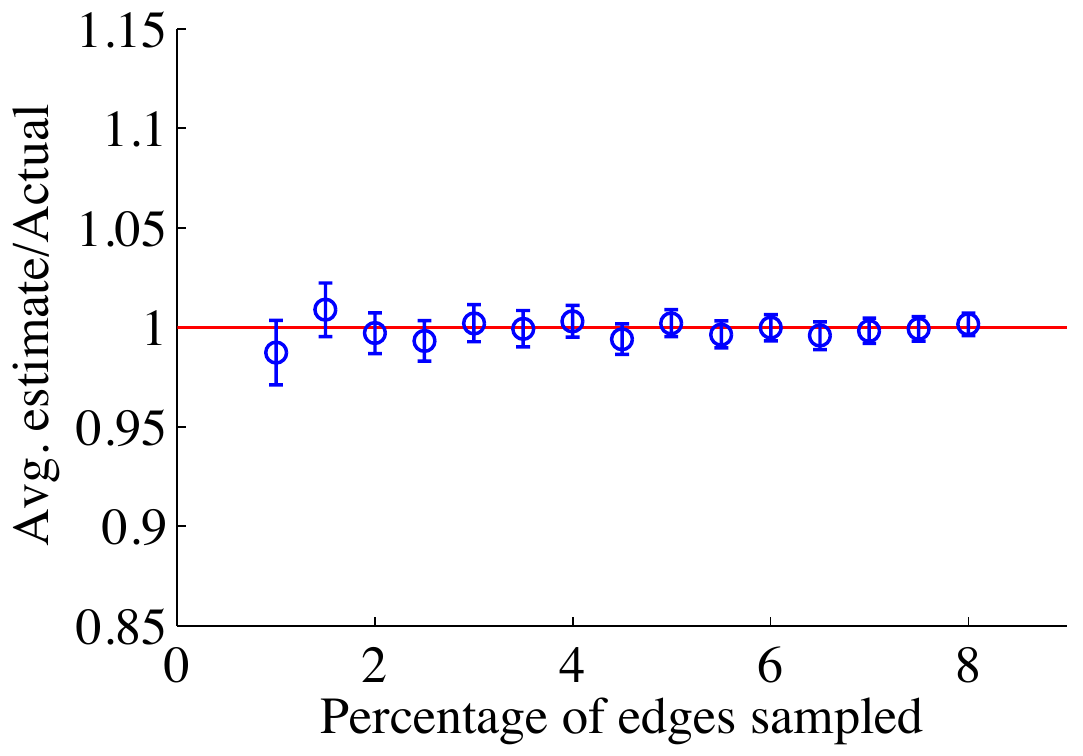}
       }
       \subfigure[{BA-5 ($\gamma=1.5$, $N_T=3,473,097$)}]{
       \includegraphics[width=2.2in]{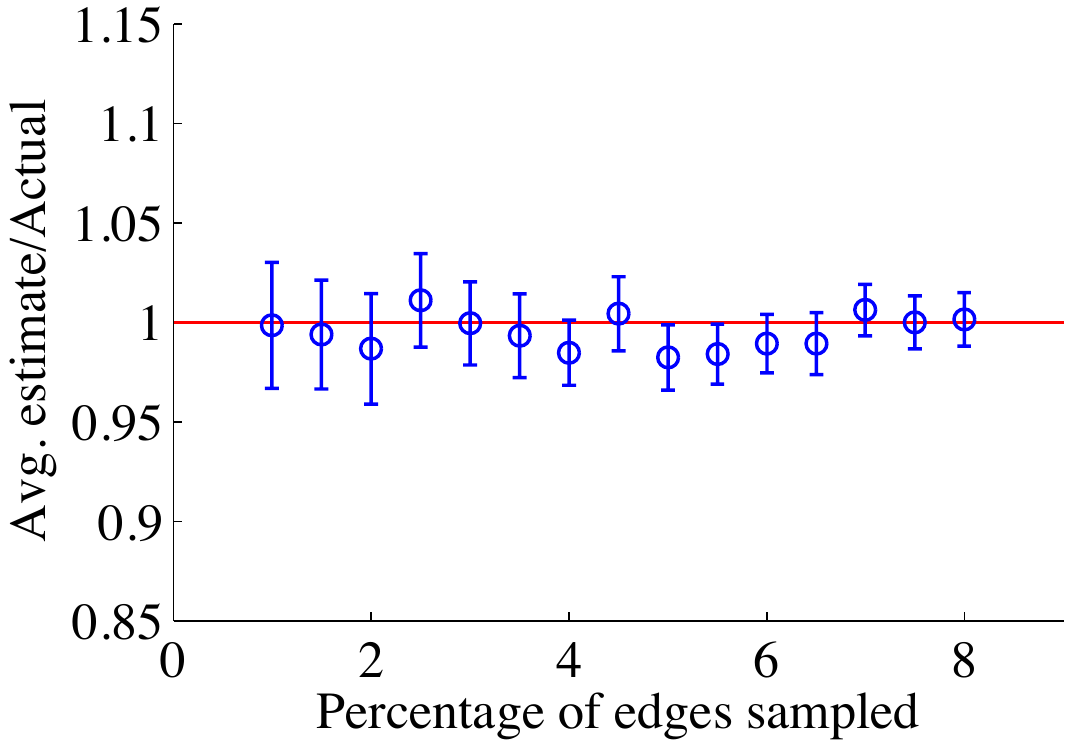}
	}
	 \subfigure[{BA-6 ($\gamma=2.0$, $N_T=3,464,420$)}]{ 
       \includegraphics[width=2.2in]{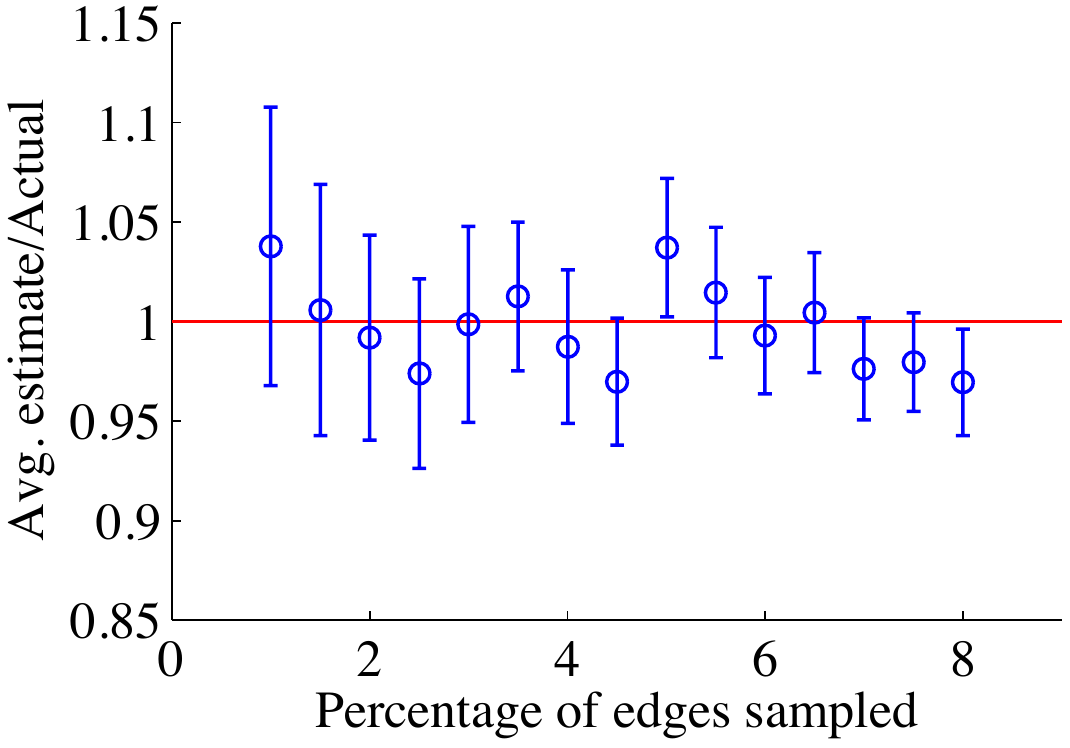}
    }
   \caption{Blue circles represent the ratio of the average estimated values of the total number of triangles to the actual value over 100 independent runs. Red line indicates 1. The blue error bars indicate 95\% confidence intervals.}\label{fig:BA_errorbar} 
\end{figure*} 

Figure \ref{fig:BA_errorbar}  shows the ratio of the average estimated
total number of triangles to the actual value for each BA graph with
increasing number of edges sampled. The error bars represent the 95\%
confidence intervals. The red line indicates 1, when the estimated and
the actual values are equal. For all of the graphs, the same sampling
fraction is used. By comparing Figures \ref{fig:BA_errorbar}$(a)$,
\ref{fig:BA_errorbar}$(b)$ and \ref{fig:BA_errorbar}$(c)$ where all of
these figures are obtained by testing on BA graphs with the same power
of the preferential attachment, we can see that the confidence
intervals are larger in the BA graphs with a smaller number of
triangles. In Figures \ref{fig:BA_errorbar}$(d)$,
\ref{fig:BA_errorbar}$(e)$ and \ref{fig:BA_errorbar}(f), the total
number of triangles in each of the tested graphs is approximately
equal but the values of $\gamma$ and the global clustering coefficient
are different. We can see that the estimate on graphs with a higher
global clustering coefficient achieves a smaller confidence interval. 

These results confirm the theoretical analysis of the relative error of the estimate presented in Section \ref{sec:estvar}. Besides the sampling fraction, the relative error of the estimate is influenced by certain properties of the graph. Our algorithm achieves better accuracy on graphs with more triangles and a higher global clustering coefficient.

\section{Conclusion}
\label{sec:conclusion}
In this paper, we propose a new framework for analyzing graphs in a dynamic system, and present an edge sampling algorithm, called {\em Edge
  Sample and Discard} (\algoname{esd}),  which estimates the total
number of triangles in a fully dynamic graph where both edge additions and
deletions are possible. With a tiny modification of the weight
parameter, \algoname{esd} can also be applied to static
graphs. 
Our algorithm achieves a significant improvement in accuracy by allowing the use of neighborhood information of the sampled edges through sending queries to the graph dataset. As illustrated in our performance analysis,
\algoname{esd} achieves much better accuracy, smaller variance, and
faster speed than the previously known state-of-the-art algorithms. In
particular, it offers a methodology to keep track of the changes in the
number of motifs of a certain type in a fully dynamic graph.


\section*{Acknowledgements}

This work was partially supported by the National Science Foundation
Award \#1250786.

%


\bibliographystyle{splncs03}
\bibliography{refs} 

\end{document}